\documentclass[12pt,prb,superscriptaddress]{revtex4}
\newcommand {\bra}[1]{\langle #1|}
\newcommand {\ket}[1]{|#1\rangle}

\newcommand {\deltav}{\mbox{\boldmath$\delta$}}
\newcommand {\omegav}{\mbox{\boldmath$\omega$}}
\newcommand {\Omegav}{\mbox{\boldmath$\Omega$}}
\newcommand {\Thetav}{\mbox{\boldmath$\Theta$}}

\newcommand {\Psiv}{\mbox{\boldmath$\Psi$}}

\usepackage{amsmath}
\usepackage{graphicx}

\begin{document}

\title{RESONANT TRANSPORT THROUGH SEMICONDUCTOR NANOSTRUCTURES} 

\author{E. R. Racec}
\affiliation{Technische Universit\"at Cottbus, Fakult\"at 1, 
             Postfach 101344, 03013 Cottbus, Germany} 
\affiliation{University of Bucharest, Faculty of Physics, PO Box MG-11,
             077125 Bucharest Magurele, Romania}
\author{P. N. Racec}
\affiliation{Technische Universit\"at Cottbus, Fakult\"at 1, 
             Postfach 101344, 03013 Cottbus, Germany}
\affiliation{National Institute of Materials Physics, PO Box MG-7,
             077125 Bucharest Magurele, Romania}
\author{Ulrich Wulf}
\affiliation{Technische Universit\"at Cottbus, Fakult\"at 1, 
             Postfach 101344, 03013 Cottbus, Germany}

\begin{abstract}
Transport through semiconductor nanostructures is a quantum-coherent
process. This review focuses on systems in which the electron's dynamics
is ballistic and the transport is dominated by the scattering from
structure boundaries. Opposite to the well-known case of the nuclear
reactions, the potentials defining semiconductor structures are
nonspherically symmetric and the asymptotic motion of the electrons is
determined by the different potential levels in the contacts.  For this
special type of potential the mathematical foundations for the scattering
theoretical description of the transport phenomena are presented. The
transport properties of the system are then derived from the scattering
matrix using the Landauer-B\"uttiker formalism. A rigorous analysis of the
analytical properties of the S matrix leads to the most general resonant
line shape described by a Fano function with a complex asymmetry
parameter. On this basis the resonant and nonresonant contributions to
the conductance and capacitance of the system are identified.
\end{abstract}

\maketitle

\section{INTRODUCTION}
\label{introduction}

A remarkable feature of the physics of semiconductor materials is the 
possibility of designing and manufacturing artificial structures in
which the electrons can be confined in zero, one, and two dimensions
\cite{ferry}. 
The region of confinement, usually called quantum system, 
is coupled to contacts through tunneling barriers which 
assure the dominance of the quantum effects in the transport phenomena.
Due to the coupling to the electrons in the Fermi sea 
of each contact the discrete electronic states of the isolated 
quantum system become resonant states \cite{kukulin}. 
They are special states of the continuum spectrum associated with 
a maximum of nonzero width of the electronic probability 
distribution density. 
The resonance width reflects the open character of the quantum system
in semiconductor nanostructures and
gives a measure of the coupling strength between the quantum 
system and contacts.
The resonances can be directly seen in the transport properties of 
nanostructures giving reason to call the transport through this special 
type of mesoscopic structures as resonant \cite{ferry}.

In 1970 Easki and Tsu \cite{esaki1} have proposed for the first time a 
semiconductor structure where electronic transport proceeds via a 
resonant tunneling process and after that in 1973 they have also proposed 
a model \cite{esaki2} to calculate the current density through  
the considered system. These were the first steps which have opened an
extremely rich field for basic and applied research. 
The resonant transport has been investigated in a multitude 
of different mesoscopic semiconductor systems\cite{ferry,kelly,ando}: 
two dimensional electron gas, quantum point contacts, 
quantum wires and quantum dots. 
Many phenomena such 
as universal conductance fluctuations \cite{univ},
the Aharonov-Bohm oscillations \cite{ahar},
the quantum Hall effect \cite{quanha},
the quantized conductance in ballistic point contacts \cite{quapoi},
Coulomb blockade oscillations \cite{coul},
chaotic dynamics in quantum dots \cite{chao},
and Kondo effect in single electron transistors \cite{kondo},
have been observed and discussed in the well-known theory pioneered by 
Landauer and B\"uttiker \cite{lanbue,bue92}.
The widely successful application of this formalism shows that electron 
transport through a mesoscopic system is quite similar to scattering 
in nuclear or atomic physics \cite{nucr}. The most important difference is 
that for transport 
the asymptotic motion of the electrons is determined by the different 
potential levels in the contacts and therefore is not free as 
in the nuclear reaction theory. In turn, the spherical symmetry of
the problem is broken and the methods to solve the scattering problem 
cannot be directly imported from the theory of nuclear reactions.
The peculiarities of the scattering potential require 
a new theoretical description of the scattering phenomena 
appropriate for the transport. 
In this theory 
not only the values of the potential in the contacts
should be taken into consideration, but also
the strength of the coupling between the quantum system and the contacts.
Recent research
on Fano resonances in the conductance of a single electron transistor 
\cite{goeres}, on fluctuations of the local density of states in the emitter
\cite{schmidt} and on Luttinger liquid behavior in ballistic transport through
quantum wires \cite{lal},  has established the importance of the 
interaction of the quantum 
system with the contacts. As an effect of this
interaction the line shape of the resonant profile is asymmetric
and cannot be described by the common Wigner-Breit distribution anymore. 
 
The main aim of this paper is to present the mathematical foundations 
for the scattering theoretical description of coherent
transport phenomena and to describe the relevant resonances for 
the transport properties. The paper is structured as follows:
Sec. \ref{st} presents the method to determine the scattering matrix 
and the scattering  functions associated to the scattering potential
of the nanostructure. In the second part of this section 
the electronic charge and current densities are deduced in the frame of the 
Landauer-B\"uttiker formalism using the second quantization technique.
Sec. \ref{res} presents an analytical theory of the quasi-isolated 
transport resonances, and the signature of these resonances in
the conductance and capacitance measurements performed on
semiconductor nanostructures is identified. The Fano functions
with a complex asymmetry parameter arise as the most general 
resonant line shape and we provide explicit expressions for
the parameters of the Fano profile.

\section{SCATTERING APPROACH TO QUANTUM TRANSPORT}
\label{st}
\subsection{THE CONSIDERED SYSTEM TYPE}
\label{systems}

The wave function which describes the electronic state of energy E 
is a solution of the Schr\"odinger equation
\begin{equation}
\left[ \frac{1}{2 m^*} \vec P^2 +V_{eff}(\vec r)-E 
\right] \Psi(E,\vec r) =0,
\label{eq-S}
\end{equation}
where $\vec P$ is the momentum operator, $V_{eff}(\vec r)$ is
the effective potential energy which is a sum of the heterojunction 
conduction band discontinuities, the electrostatic potential 
due to the ionized donors and acceptors,
the self-consistent Hartree and exchange potentials due to free 
carriers and external potentials. 
The electronic states in mesoscopic systems are easily described
within the effective mass approximation whose validity requires
that the envelope function $\Psi(E,\vec r)$ be slowly varying over
dimensions comparable to the unit cell of the crystal \cite{bastard}.

\begin{figure}[t]
\begin{center}
\vspace*{-0.25cm}
\noindent\includegraphics[height=2.25in]{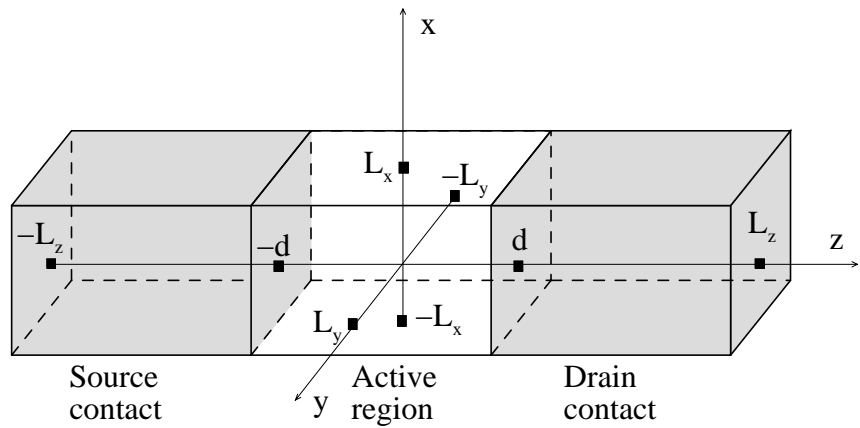}
\caption{A system relevant for transport properties: active region 
         with size $2d$ sandwiched between source and drain contacts. 
         The dimensions of the system are large enough to be extended 
         to infinity ($d \ll L_x,L_y,L_z$). }
\label{skizze}
\end{center}
\end{figure}

We restrict our analysis to the case 
of semiconductor nanostructures presented schematically 
in Fig.  \ref{skizze}, in which the current flows
only in one direction, from the source- to the drain contact,
called transport direction; any kind of 
residual current is neglected. 
This means that  the electrons are either bound
in the lateral directions or freely moving in a plane 
perpendicular to the transport direction, so that the 
mean value of the lateral current is zero. 
Consequently, the potential energy in which the electrons
move should have a perpendicular component $V_{\perp}(\vec r)$,
which we assume independent of the coordinate in the transport direction,
i.e. $V_{\perp}(\vec r) \equiv V_{\perp}(\vec r_{\perp})$.
For the structures which contain only a 2DEG and no lateral modulation 
 $V_{\perp}(\vec r_{\perp})=0$,
whereas for  quantum dots $V_{\perp}(\vec r_{\perp})$
is considered either a parabolic 
or a square infinite well potential energy. 
We assume, also, that except for this overall component 
$V_{\perp}(\vec r_{\perp})$, there is no other lateral potential
which can supplementary confine the electrons in  a  certain region 
of the structure.
Thus, the potential energy is separable,
\begin{equation}
V_{eff}(\vec r) = V(z) + V_{\perp}(\vec r_{\perp}).
\label{pot}
\end{equation}
This special form of $V_{eff}(\vec r)$ seems to be a strong restriction
and it is not easy to imagine its validity for
structures with complex geometry, such as a single electron
transistor \cite{goeres}, 
but it is the price payed for a 
good analytical description of the transport phenomena 
in semiconductor nanostructures.

Due to the separable form of the potential, a particular solution
of the Schr\"odinger equation (\ref{eq-S}) associated to the energy 
$E=E_{\perp}+\epsilon$ is
\begin{equation}
\Psi^{(s)}_{\nu}(E;\vec r)
   = \phi_{\nu}(\vec r_{\perp})
     \psi^{(s)}(\epsilon,z),
\label{Psi-fq}
\end{equation}
where $\nu$ indexes the energy levels $E_{\perp}^{\nu}$
associated with the 
motion in the lateral directions.
Here $s$ is the degeneracy index for the electron motion 
in the transport direction.
The functions $\phi_{\nu}(\vec r_{\perp})$ and the energies
$E_{\perp}^{\nu}$ are the solutions of the eigenvalue problem
\begin{equation}
\left[ \frac{1}{2 m^*} \vec P_{\perp}^2 +V_{\perp}(\vec r_{\perp})
- E_{\perp}^{\nu}
\right] \phi_{\nu}(\vec r_{\perp}) =0.
\label{eq-S-2D}
\end{equation}
The Hamilton operator of this problem is a Hermitian one and,
consequently, the eigenfunction set 
$\left\{\phi_{\nu}(\vec r_{\perp})\right\}_{\nu}$
forms a basis for which the orthogonality and 
completeness relations
\begin{eqnarray}
\int d \vec r_{\perp} \, \phi^*_{\nu}(\vec r_{\perp})
                        \phi_{\nu'}(\vec r_{\perp})
& = & \delta_{\nu \nu'}, 
\nonumber
\\
\sum_{\nu} \phi^*_{\nu}(\vec r_{\perp})
           \phi_{\nu}(\vec r^{\,'}_{\perp})
& = & \delta(\vec r_{\perp}-\vec r^{\,'}_{\perp}),
\label{basis-2D}
\end{eqnarray}
are valid. The index $\nu$ is generic;
for a free electron system in a plane perpendicular to the 
transport direction this index should be replaced by a 
2D wave vector $\vec k_{\perp}$. 
The energy spectrum becomes  continuous
($E_{\perp} = \hbar^2 k_{\perp}^2/2 m^*$).
Thus, in Eqs. (\ref{basis-2D}) 
$\delta_{\nu \nu'}$ is transformed into 
$\delta(\vec k_{\perp}-\vec k'_{\perp})$
and the sum over $\nu$ becomes an integral over $\vec k_{\perp}$
($\sum_\nu \rightarrow A/(2 \pi)^2 \int d \vec k_\perp$, 
where $A$ is the lateral area of the system). 
Every energy level $E_{\perp}$ is degenerate
with an infinite degeneracy. In the opposite limit, 
for a complete confinement of the system in the lateral directions, 
the energy levels are discrete and nondegenerate, and $\nu$ is
substituted by a quantum number.

The $z$-dependent function in Eq. (\ref{Psi-fq})
is a solution
to the one dimensional eigenvalue problem
\begin{equation}
\left[ \frac{1}{2 m^*} P_z^2 +V(z) - \epsilon
\right] \psi^{(s)}(\epsilon,z) = 0.
\label{eq-S-1D}
\end{equation}
For nanostructures, 
the potential energy $V(z)$ has some common properties:
There is an active region 
which is small at the scale of the whole system.
Inside this region, the electrons are elastically scattered
from interfaces between different layers or between allowed and not 
allowed (depletion) domains in the case of a structured 2DEG.
This region is also called {\it scattering region}
and it is embedded between two contacts, 
which are practically semi-infinite homogeneous semiconductors.
The electrons inside the contact regions 
move in a potential energy given only
by the bulk conduction band structure.
A relation between the surfaces of the scattering region and
the interfaces between heterostructure layers or domains of the 2DEG
can not be established.
Each interface can cause elastic scattering 
and should be included in the scattering region. Thus, the planes
$z=\pm d$ are chosen inside the homogeneous materials of the contacts,
far enough from any interface and, of course, the choice is not unique.  
From the mathematical
point of view, the potential energy $V(z)$ is a
function which varies considerably over small distances
inside the {\it scattering region}
and is constant outside, i.e. $V(z < -d) = V_1$ and
$V(z > d) = V_2$, where $\pm d$ are the limits of the scattering
region. Further we introduce the notations: $V_m=\min(V_1,V_2)$
and $V_M=\max(V_1,V_2)$. 

\subsection{S-MATRIX AND SCATTERING STATES}
\label{smat}

The above considerations about the potential energy experienced by  the
electrons allow us to reduce the Schr\"odinger equation (\ref{eq-S})
to the effectively one dimensional eigenvalue problem
(\ref{eq-S-1D}). This problem is essentially a scattering problem for
a particle with mass $m^*$, in a potential with
nonspherical symmetry, and it is usually  solved in the scattering matrix 
approach \cite{lanbue}. 

In the stationary description of the scattering process
it is assumed that the wave function of the system at
large distances  from the scattering region is a superposition
of an incident- and a scattered wave \cite{taylor}.
In our case the electrons are elastically scattered
inside the domain $[-d,\,d]$ 
and for all points outside this region
the $z$-dependent part of the wave function is
given by
\begin{equation}
\psi(\epsilon,z) = \left\{
                   \begin{array}{ll}
                     \psi^{in}(\epsilon,-d) \exp[i k_1 (z+d)]
                    +\psi^{out}(\epsilon,-d) \exp[-i k_1 (z+d)],
                    & z \le -d \\
                    & \\
                     \psi^{in}(\epsilon,d) \exp[-i k_2 (z-d)]
                    +\psi^{out}(\epsilon,d) \exp[ i k_2 (z-d)], 
                    & z \ge d
                   \end{array}
                   \right.
\label{psi-out}
\end{equation}             
where $\psi^{in/out}(\epsilon,\pm d)$ are complex coefficients and
\begin{equation}
k_s(\epsilon) = \sqrt{\frac{2 m^*}{\hbar^2} (\epsilon - V_s)},
\qquad s =1,2.
\label{k}
\end{equation}
If $\epsilon > V_s$, then $k_s$ is a nonzero positive number according
to the definition of the real square root function.
However, if $\epsilon < V_s$, we have to take the first branch of the 
complex square root function, so that $k_s = i \left| k_s \right|$.
At this stage 
we would like to point out the main difference between a spherically
symmetric scattering problem like in the nuclear physics and a transport
scattering problem: in the former only one wave vector
for every value of the total energy is defined, 
while in the latter there are two wave
vectors, $k_s$, $s=1,2$, for each energy associated with the electron
motion in the transport direction.
If the materials of the source and drain contacts are different
or when a current flows through the structure,
$V_1$ and $V_2$ are different and consequently $k_1 \neq k_2$.

For a fixed energy  $\epsilon$ there are at most two independent solutions
of Eq. (\ref{eq-S-1D}) and, consequently, at most two 
expansion coefficients $\psi^{in/out}(\epsilon,\pm d)$ 
can be independent.
We choose them $\psi^{in}(\epsilon,\pm d)$. The coefficients 
 corresponding to the outgoing waves in Eq.  (\ref{psi-out}) 
should be expressed in terms of
$\psi^{in}(\epsilon,\pm d)$, 
\begin{equation}
\left(
\begin{array}{l}
          \psi^{out}(\epsilon,-d) \\
          \psi^{out}(\epsilon,+d)
\end{array}
\right)
=
{\bf S}(\epsilon)
\left(
\begin{array}{l}
         \psi^{in}(\epsilon,-d) \\
         \psi^{in}(\epsilon,+d)  
\end{array}
\right).
\label{S-mat}
\end{equation}
The matrix ${\bf S}(\epsilon)$, defined for every value of the energy
$\epsilon$ associated with the motion of the electron in the transport 
direction, contains all the information
about the collisions in the system \cite{taylor} 
and is called {\it scattering matrix}.
As a consequence of the existence of two wave vectors for each value of
the energy $\epsilon$ the scattering processes are described by a 
$2 \times 2$ matrix even in the simplest case. 

For energies smaller than the absolute minimum of the potential 
energy,
the solution given by Eq. (\ref{psi-out}) should be zero
and $\psi^{in/out}(\epsilon,\pm d) = 0$.
If the energies are greater than $\min[V(z)]$,
the eigenvalue spectrum of the one dimensional
Schr\"odinger equation (\ref{eq-S-1D})
has a discrete part for $V_m > \epsilon > \min[V(z)]$
 associated with bound states
and a continuous part for $\epsilon > V_m$
corresponding to the scattering states.
We are interested in describing
elastic processes which contribute to the transport phenomena  and
in this case the bound states do not play any role;
Further we assume $V(z) \ge V_m$.
The continuous part of the energy spectrum has 
two important regions: $\epsilon \ge V_M$,
for which two independent solutions
of Eq. (\ref{eq-S-1D}) exist, and $V_m \le \epsilon \le V_M$  
with only one solution.
For the degenerate energy levels there is not an unique way of
defining the eigenfunctions 
and we further prefer to use the so called {\it scattering 
functions} \cite{lanbue},
\begin{eqnarray} 
\psi^{(1)}(\epsilon,z)&=&\frac{\theta(\epsilon-V_1)}{\sqrt{2 \pi}}
                         \left\{
                         \begin{array}{ll} 
			         \exp{[i k_1 (z+d)]}
                                +{\bf S}_{11}(\epsilon) \exp{[-i k_1 (z+d)]},
                                & z \le -d \\
                                & \\
                                {\bf S}_{21}(\epsilon) \exp{[i k_2 (z-d)]},
                                & z \ge d 
			\end{array}
                        \right.,
\label{psi-out-1} \\
\nonumber \\
\psi^{(2)}(\epsilon,z)&=&\frac{\theta(\epsilon-V_2)}{\sqrt{2 \pi}}
                         \left\{
                         \begin{array}{ll}
			          {\bf S}_{12}(\epsilon) \exp{[-i k_1 (z+d)]},
                                & z \le -d \\
                                & \\
                                 \exp{[-i k_2 (z-d)]}
                                +{\bf S}_{22}(\epsilon) \exp{[i k_2 (z-d)]},
                                & z \ge d 
			\end{array}
                        \right..
\label{psi-out-2}
\end{eqnarray}
$\psi^{(1)}(\epsilon,z)$
corresponds to a particle which comes into the scattering system
from the left reservoir ($s=1$)
and is either transmitted or reflected.  
We can identify the element ${\bf S}_{11}(\epsilon)$ of the
{\bf S} matrix with the reflection amplitude, $r^{(1)}(\epsilon)$,
and the element ${\bf S}_{21}(\epsilon)$ with the transmission amplitude,
$t^{(1)}(\epsilon)$.
$\psi^{(2)}(\epsilon,z)$
corresponds to a particle coming from the
right side of the system ($s=2$).
The reflection and transmission amplitudes for this particle are
$r^{(2)}(\epsilon) = {\bf S}_{22}(\epsilon)$
and $t^{(2)}(\epsilon) = {\bf S}_{12}(\epsilon)$, respectively.
In the case $\epsilon < V_s$, the time translation invariance requires
that $|r^{(s)}(\epsilon)|=1$, $s=1,2$.

The step functions
\begin{equation}
\theta(\epsilon-V_s) = \begin{cases}
                          1, & \epsilon > V_s \\
                          0, & \epsilon < V_s 
		       \end{cases},
\qquad s=1,2,
\label{theta}
\end{equation}
ensure that for $V_m < \epsilon \le V_M$
there is only one scattering function for each energy $\epsilon$.

Generally, the wave functions are solutions of a differential equation
and are defined up to a constant factor. In the above expression,
we fix this constant, so that, in the limit of free particles
[$V(z) \rightarrow 0$, $d \rightarrow 0$],
the scattering functions [Eqs. (\ref{psi-out-1}-\ref{psi-out-2})] 
become eigenfunctions 
of the momentum operator $P_z$. 

The matrix elements of ${\bf S}$, i.e. the transmission and reflection
coefficients,  are formally determined by the continuity conditions
of the scattering  functions and their first derivatives at $z=\pm d$,
\begin{equation}
\Psiv(\epsilon) = \frac{1}{\sqrt{2\pi}} 
                       \Thetav(\epsilon)
                       \left[ {\bf 1} + {\bf S}(\epsilon) \right]
\label{Psi-mat}
\end{equation}
with the matrices 
$\left(\Psiv(\epsilon)\right)_{ss'} = \Psi^{(s')}(\epsilon;(-1)^s d)$
and $\Thetav_{ss'}=\theta(\epsilon-V_s) \delta_{ss'}$, $s,s'=1,2$,
and
\begin{equation}
\Psiv_S(\epsilon) =-\frac{i}{\sqrt{2\pi}}
                         \frac{\pi}{2d} 
                         \Thetav(\epsilon) {\bf K}(\epsilon)
                         \left[ {\bf 1} - {\bf S}(\epsilon) \right]
\label{Psi-S-mat}
\end{equation}
with $\left(\Psiv_S(\epsilon)\right)_{ss'} 
= (-1)^s \left. \frac{\partial \Psi^{(s')}}
                     {\partial z}
         \right|_{z=(-1)^s d}$ 
and ${\bf K}_{ss'} = k_s/(\pi/2d) \delta_{ss'}$,
$s,s'=1,2$, respectively.
We can use these conditions only after the calculation of the scattering
functions inside the scattering region.

In case of a general potential inside the scattering region,
we can not represent the 
functions $\psi^{(s)}(\epsilon,z)$
by elementary functions between $-d$ and $d$ and it is necessary to
expand them into a basis of eigenfunctions of a solvable  problem. 
In the {\it R matrix formalism} 
the Wigner-Eisenbud basis is considered,
i.e. the eigenbasis of the Hamilton operator corresponding
to the closed counterpart of the studied scattering problem,
\begin{equation}
\left[ -\frac{\hbar^2}{2 m^*} \frac{d^2}{d z^2}
       +V(z)-\epsilon_l
\right] \chi_l(z) = 0
\label{WE-eq}
\end{equation}
with the boundary conditions
$ \left . d \; \chi_l/dz \right|_{z = \pm d} = 0$.
$\chi_l(z)$ are real functions defined only for $z \in [-d,d]$
and verify the orthogonality and the completeness
relations,
\begin{eqnarray}
\int_{-d}^{d} dz \; \chi_l(z) \chi_{l'}(z) = \delta_{ll'},
\label{WE-ortho}
\\
\sum_{l=1}^{\infty} \chi_l(z) \chi_{l}(z') = \delta(z-z'),
\label{WE-complete}
\end{eqnarray}
respectively. 
The R matrix  method developed by Wigner and Eisenbud \cite{WE}
and intensively used in the nuclear reaction theory 
\cite{lane},
has become important in the last ten years 
also for describing transport phenomena 
\cite{Rmat-transport}.

We expand the wave functions inside the scattering region, $z \in [-d,d]$,
in terms of the Wigner-Eisenbud 
functions 
\begin{equation}
\psi^{(s)}(\epsilon,z) = \sum_{l=1}^{\infty} 
                           a_l^{(s)}(\epsilon) \chi_l(z)
\label{psi-in}
\end{equation}
and the expansion coefficients $a_l^{(s)}$ are given by
\begin{equation}
a_l^{(s)}(\epsilon) = \int_{-d}^{d} dz \; 
                        \chi_l(z) \psi^{(s)}(\epsilon,z)
\label{coeff-psi}
\end{equation}
for each $s=1,2$ and $l \ge 1$. 
To determine 
$a^{(s)}(\epsilon)$ 
we use the Schr\"odinger equation (\ref{eq-S-1D})
and Eq. (\ref{WE-eq}) satisfied by the Wigner-Eisenbud functions.
We multiply them
by $\chi_l(z)$ and $\psi^{(s)}(\epsilon,z)$, respectively,
and the difference between the resulting equations is integrated over 
the interval $[-d,d]$.
Performing an integration by parts in the kinetic
energy term 
and identifying the integral 
on the right side with the coefficient $a^{(s)}_l$,
defined by (\ref{coeff-psi}), we obtain
\begin{equation}
a_l^{(s)}(\epsilon) = \frac{\hbar^2}{2 m^*}
                      \frac{\chi_l(-d)}{\epsilon-\epsilon_l}
		      \left. \frac{\partial \psi^{(s)}}{\partial z}
		      \right|_{z=-d}
                     -\frac{\hbar^2}{2 m^*}
                      \frac{\chi_l(d)}{\epsilon-\epsilon_l}
		      \left. \frac{\partial \psi^{(s)}}{\partial z}
	              \right|_{z=d}.
\label{coeff-psi-2}
\end{equation}
Inserting Eq. (\ref{coeff-psi-2})  
into Eq. (\ref{psi-in})
we obtain the expression of the scattering functions 
between $-d$ and $d$
as a function of their derivatives at the edges of the
scattering area, 
\begin{equation}
\psi^{(s)}(\epsilon,z)=\frac{2d}{\pi}
                       \left[ R(\epsilon;-d,z)
                              \left. \frac{\partial \psi^{(s)}}
                                          {\partial z}
                              \right|_{z=-d}
                             -R(\epsilon;d,z)
                              \left. \frac{\partial \psi^{(s)}}
                                          {\partial z}
                              \right|_{z=d}
                       \right],
\label{psi-in-II}
\end{equation}
where the R function is defined as
\begin{equation}
R(\epsilon;z,z') = \frac{\hbar^2}{2 m^*}
                   \frac{\pi}{2d}
                   \sum_{l=1}^{\infty}
                      \frac{\chi_l(z) \; \chi_l(z') }
                           {\epsilon-\epsilon_l}.
\label{R-def}
\end{equation}
The above expression of the R function has the advantage
of being dimensionless.
Together with the continuity condition (\ref{Psi-S-mat})
the relation (\ref{psi-in-II}) gives the scattering functions inside the
scattering region ($-d \le z \le d$) in terms of the ${\bf S}$ matrix,
\begin{eqnarray}
\psi^{(1)}(\epsilon,z)&=&\frac{\theta(\epsilon-V_1)}{\sqrt{2\pi}}
                         \frac{2d}{\pi}
                          \left\{
                            i k_1 \left[1 - {\bf S}_{11}(\epsilon)
                                  \right] R(\epsilon;-d,z) 
                           -i k_2 {\bf S}_{21}(\epsilon)
                            R(\epsilon;d,z) 
                         \right\},
\label{psi-in-1} \\
\psi^{(2)}(\epsilon,z)&=&\frac{\theta(\epsilon-V_2)}{\sqrt{2\pi}}
                          \frac{2d}{\pi}
                          \left\{
                           -i k_1 {\bf S}_{12}(\epsilon)
                            R(\epsilon;-d,z) 
                           +i k_2 \left[1 - {\bf S}_{22}(\epsilon)
                                  \right] R(\epsilon;d,z)
                          \right\}.
\label{psi-in-2}
\end{eqnarray}
The values of the scattering
functions on the surface of the scattering domain
are then given by
\begin{equation}
\Psiv(\epsilon)
 =\frac{i}{\sqrt{2 \pi}} 
   \Thetav(\epsilon)
   {\bf R}(\epsilon) {\bf K}(\epsilon)
   \left[ {\bf 1} - {\bf S}(\epsilon)
   \right],
\label{Psi-mat-2}
\end{equation}
with the ${\bf R}$ matrix defined as 
\begin{equation}
{\bf R}(\epsilon) = \begin{pmatrix}
                              R(\epsilon;-d,-d) & R(\epsilon; d,-d) \\
                              R(\epsilon;-d, d) & R(\epsilon; d,d)
                     \end{pmatrix}.
\label{R-mat-def}
\end{equation} 

The scattering functions are continuous at the edges of the 
scattering system 
and we exploit this condition to determine the relation between 
the ${\bf R}$ and the ${\bf S}$ matrix.
Equating the two expressions (\ref{Psi-mat}) and (\ref{Psi-mat-2})
of the matrix $\Psiv$ we infer  
\begin{equation}
{\bf S}(\epsilon) = {\bf 1} 
                   - 2 \left[ {\bf 1} + i {\bf R}(\epsilon) {\bf K}(\epsilon) 
                       \right]^{-1}
\label{rel-SR}
\end{equation}
for each value of the energy $\epsilon$ in the domain of the scattering
states. For $V_m < \epsilon  < V_M$
the above relation can only be used to calculate the matrix elements 
${\bf S}_{1s}$ and ${\bf S}_{2s}$ with $s$ satisfying the condition
$\epsilon > V_s$, $s=1,2$.
Using Eq. (\ref{rel-SR}) the elements of the scattering matrix can be
effectively determined in terms of Wigner-Eisenbud functions and energies.
Furthermore, the scattering functions, Eqs.
(\ref{psi-out-1}-\ref{psi-out-2}) and 
(\ref{psi-in-1}-\ref{psi-in-2}),
can be evaluated in every point of the system,
even inside the scattering system.

As shown in the first part of this section,
the elements of the ${\bf S}$ matrix defined by Eq. (\ref{S-mat}) 
are in fact the reflection and transmission amplitudes 
for the particles coming from the left and from the right side of the system.
However, for describing transport phenomena in semiconductors
in the frame of the Landauer-B\"uttiker formalism \cite{lanbue},
the reflection and transmission probabilities, 
$R^{(1)}(\epsilon)= \left| r^{(1)}(\epsilon) \right|^2$,
$R^{(2)}(\epsilon)= \left| r^{(2)}(\epsilon) \right|^2$,
$T^{(1)}(\epsilon)= \left| t^{(1)}(\epsilon) \right|^2$
and
$T^{(2)}(\epsilon)= \left| t^{(2)}(\epsilon) \right|^2$,
play a central role. We introduce further 
the current scattering matrix \cite{Rmat-transport}
\begin{equation}
\tilde{\bf S}(\epsilon) = {\bf K}^{1/2}(\epsilon)
                          {\bf S }(\epsilon)
                          {\bf K}^{-1/2}(\epsilon),
\label{St-mat}
\end{equation}
having the property that
$\left| \tilde{\bf S}_{11}(\epsilon) \right|^2 = R^{(1)}(\epsilon)$, 
$\left| \tilde{\bf S}_{12}(\epsilon) \right|^2 = T^{(2)}(\epsilon)$, 
$\left| \tilde{\bf S}_{21}(\epsilon) \right|^2 = T^{(1)}(\epsilon)$, and
$\left| \tilde{\bf S}_{22}(\epsilon) \right|^2 = R^{(2)}(\epsilon)$.
The relation (\ref{rel-SR}) between the ${\bf R}$ and the ${\bf S}$ matrix
can be rewritten into an equivalent form
\begin{equation}
\tilde{\bf S}(\epsilon) = {\bf 1} - 2 \left[ {\bf 1}
                                            +i\Omegav(\epsilon)
                                      \right]^{-1},
\label{rel-StR}
\end{equation}
where the matrix $\Omegav$ of rank two is defined as
\begin{equation}
\Omegav(\epsilon) = {\bf K}^{1/2}(\epsilon)
                        {\bf R}(\epsilon)
                        {\bf K}^{1/2}(\epsilon)
                      = \sum_{l=1}^{\infty} 
                         \frac{\omegav_l(\epsilon) }
                           {\epsilon-\epsilon_l},
\label{Omega}
\end{equation}
with
\begin{equation}
\left( \omegav_l(\epsilon) \right)_{ss'}
= \frac{\hbar^2}{2 m^*}
  \left(k_s k_{s'} \right)^{1/2}
  \chi_l((-1)^{s} d) \, \chi_l((-1)^{s'} d),
\qquad s,\,s'=1,\,2
\label{omega}
\end{equation}
for all $l \ge 1$. Per construction $\Omegav$ is a symmetrical matrix
with real elements in the domain of the scattering states
with $\epsilon > V_M$.
This property leads to the unitarity of 
the current scattering matrix,
\begin{equation}
 \tilde{\bf S}(\epsilon) \tilde{\bf S}^{\dagger}(\epsilon)
=\tilde{\bf S}^{\dagger}(\epsilon) \tilde{\bf S}(\epsilon)
={\bf 1},
\qquad \epsilon \ge V_M,
\label{unitarity}
\end{equation}
which includes the reciprocity relations,
$R^{(1)}(\epsilon) = R^{(2)}(\epsilon)  =  R(\epsilon)$, 
$T^{(1)}(\epsilon) = T^{(2)}(\epsilon)  =  T(\epsilon)$,
and the well-known relation of the flux conservation,
$R(\epsilon) + T(\epsilon) = 1$.

The scattering functions 
$\psi^{(s)}(\epsilon,z)$ 
form an orthogonal and complete system
in the case of a potential energy $V(z)$ which does not allow bound states
\cite{bue92}.
They are eigenfunctions 
of the Hamilton operator [Eq. (\ref{eq-S-1D})]
and the self-adjointness of this operator \cite{thesis} 
allows us to consider the set as complete,
\begin{equation}
\sum_{s=1,2} \int_{V_s}^{\infty} d\epsilon \; g_s(\epsilon) 
\left( \psi^{(s)}(\epsilon;z) \right)^{*} \psi^{(s)}(\epsilon;z')
= \delta(z - z'),
\label{complete}
\end{equation}
where 
\begin{equation}
g_s(\epsilon)= \frac{m^*}{\hbar^2 k_s(\epsilon)}, 
\qquad s=1,2
\label{gs}
\end{equation}
defines the 1D density of states.
The orthogonality condition is written as
\begin{equation}
\int_{-\infty}^{\infty} d z
\left( \psi^{(s)}(\epsilon,z) \right)^{*} \psi^{(s')}(\epsilon',z)
= \theta(\epsilon-V_s) \delta_{ss'} \delta(\epsilon-\epsilon')/g_s(\epsilon)  
\label{ortho}
\end{equation}
as demonstrated in Appendix \ref{A-ortho}.

\subsection{OBSERVABLES}

The conduction electrons in a semiconductor are considered 
as an electron gas embedded in a positively charged medium 
which maintains the overall charge neutrality of the system.
The simplest approximation for the description of an electron gas
is to neglect all interactions: the Coulomb interaction of the 
electrons with each other and the interaction of the electrons
with the positive background. Every electron is then independent from 
 other electrons and subject only to external forces.  
Each state of the system is described by the field operators
$\hat \Psi(\vec r)$ and $\hat \Psi^{\dagger}(\vec r)$.
Using the eigenbasis of the one-particle Hamiltonian
associated with the Schr\"odinger equation (\ref{eq-S}), 
we can represent $\hat\Psi(\vec r)$ and 
$\hat \Psi^{\dagger}(\vec r)$ in terms of
creation and destruction operators for the states $(\nu,s,\epsilon)$
of the field, $c_{\nu s}(\epsilon)$ and 
$c^{\dagger}_{\nu s}(\epsilon)$,
respectively. Thus, 
\begin{eqnarray}
\hat \Psi(\vec r)&=&\sum_{\nu,s}  
                    \int d \epsilon  g_s(\epsilon)
                       \phi_\nu(\vec r_{\perp})
                       \psi^{(s)}(\epsilon,z) 
                       c_{\nu s}(\epsilon),
\label{Psi}
\\
\hat \Psi^{\dagger}(\vec r)&=&\sum_{\nu,s} 
                              \int d \epsilon g_s(\epsilon) 
                              \phi^*_{\nu}(\vec r_{\perp})
                              \left( \psi^{(s)}(\epsilon,z) 
                              \right)^*
                              c_{\nu s}^{\dagger}(\epsilon),
\label{Psi+}
\end{eqnarray}
with the 1D density of states $g_s(\epsilon)$, $s=1,2$ defined by 
Eq. (\ref{gs}).
The operators $\hat \Psi(\vec r)$ and $\hat \Psi^{\dagger}(\vec r)$
satisfy the fermion type anticommutation relations; consequently the
creation and destruction operators have the 
following properties \cite{bue92},
\begin{eqnarray}
\left\{ c_{\nu s}(\epsilon), \,
        c_{\nu' s'}^{\dagger}(\epsilon')
\right\} 
&= & \theta(\epsilon-V_s) 
     \delta_{\nu\nu'} \delta_{ss'} 
     \delta(\epsilon-\epsilon')/g_s(\epsilon), 
\nonumber \\
\left\{ c_{\nu s}(\epsilon), \,
        c_{\nu' s'}(\epsilon')
\right\}& = & 0, 
\label{anticomm}
\\
\left\{ c_{\nu s}^{\dagger}(\epsilon), \,
        c_{\nu' s'}^{\dagger}(\epsilon')
\right\} &= & 0.
\nonumber
\end{eqnarray}

The many-particle Hamiltonian of the electron system without mutual
interaction, 
\begin{equation}
\hat H =  \int d \vec r \; \hat \Psi^{\dagger}(\vec r)
                           \left[ \frac{1}{2 m^*} \vec P^2
                                 +V_{\perp}(\vec r_{\perp})
                                 +V(z)
                           \right]
                           \hat \Psi(\vec r),
\end{equation}
can also be written in terms of creation and destruction operators,
\begin{equation}
\hat H = \sum_{\nu,s}  
         \int d \epsilon \; g_s(\epsilon)
              E_{\nu}(\epsilon) 
              c_{\nu s}^{\dagger}(\epsilon)
              c_{\nu s}(\epsilon),
\label{Ham}
\end{equation}
where $ E_{\nu}(\epsilon) = E_{\perp}^{\nu} +\epsilon$
is the energy of the single particle state $(\nu,s,\epsilon)$.
An eigenvector $\ket{\alpha}$ of the Hamiltonian (\ref{Ham}) 
is  completely determined
by the occupation numbers $n^{(\alpha)}_{\nu s}(\epsilon)$
of every single particle state $(\nu,s,\epsilon)$,
\begin{equation}
\ket{\alpha} = \underset{(\nu,s,\epsilon)}{\bigotimes}
               \ket{n^{(\alpha)}_{\nu s}(\epsilon)}
\label{alpha-state}
\end{equation}
with $c_{\nu s}^{\dagger}(\epsilon) c_{\nu s}(\epsilon)
      \ket{n^{(\alpha)}_{\nu s}(\epsilon)}
      = n^{(\alpha)}_{\nu s}(\epsilon)
        \ket{n^{(\alpha)}_{\nu s}(\epsilon)}$.
According to the Pauli principle, at most one particle may
occupy any fermion state so that these occupation numbers 
are restricted to the values 0 and 1.

The eigenvectors of the Hamiltonian, $\ket{\alpha}$,
form an orthonormal and complete system of vectors
in the Hilbert space and each general state of the isolated electron gas 
can be given as a
superposition of the pure states of $\hat H$.
According to the postulates of the statistical mechanics we suppose
this superposition of pure states as incoherent \cite{huang},
so that the mean values of the electron charge 
and current densities in a mixed state are given by
\begin{equation}
q(\vec r) =-2e \mbox{Tr}\left[ \hat \rho \,
                               \hat \Psi^{\dagger}(\vec r) \hat \Psi(\vec r)
                        \right] 
\label{qq}
\end{equation}
and 
\begin{equation}
\vec j(\vec r) = \frac{2e \hbar}{m^*} 
                 \mbox{Tr}\left\{ \hat \rho \,
                                  \mbox{Im} \left[ \hat \Psi^{\dagger}(\vec r)
                                                   \nabla \hat \Psi(\vec r)
                                            \right]
                          \right\},
\label{jj}
\end{equation}
respectively,
where $\hat \rho$ is the density matrix,
$e=|e|$ is the elementary charge
and the factor 2 comes from the spin degeneracy.
In the definition of the electronic current density we have taken into
account the conventional sign of the current.

In the diagonal representation of the Hamiltonian  
the density matrix of the isolated system is also diagonal \cite{huang},
$\hat \rho_{\alpha \alpha'} = p_\alpha \delta_{\alpha \alpha'}$,
where $p_\alpha$ denotes here the probability of the system 
for being in the pure state $\ket{\alpha}$ and
satisfies the relation $\sum_\alpha p_\alpha=\mbox{Tr}[\hat \rho]=1$.
Inserting Eqs. (\ref{Psi}) and (\ref{Psi+})
of the field operators in definitions (\ref{qq}) and (\ref{jj})
matrix elements such as 
$\bra{\alpha} c^\dagger_{\nu s}(\epsilon) c_{\nu' s'}(\epsilon')
\ket{\alpha}$ occur. The special form of the orthogonality 
condition for the scattering states [Eq. (\ref{ortho})] yields
\begin{equation}
\bra{\alpha} 
c^\dagger_{\nu s}(\epsilon) c_{\nu' s'}(\epsilon') 
\ket{\alpha} 
= n^{(\alpha)}_{\nu s}(\epsilon) 
  \theta(\epsilon-V_s) \delta_{\nu \nu'} \delta_{ss'} 
  \delta(\epsilon-\epsilon') / g_s(\epsilon).
\end{equation}
Therefore, the electron charge and
current densities become
\begin{equation}
q(\vec r)  =-2e\sum_{\nu,s} 
             \int_{V_s}^{\infty} d \epsilon \; g_s(\epsilon)
             \left| \phi_{\nu}(\vec r_{\perp}) \right|^2
             \left| \psi^{(s)}(\epsilon,z) \right|^2
             \bar{n}_{\nu s}(\epsilon)
\label{qq-2}
\end{equation}
and
\begin{eqnarray}
\vec j(\vec r)&=& \frac{2e \hbar}{m^*}
                  \sum_{\nu,s}
                  \int_{V_s}^{\infty} d \epsilon \; g_s(\epsilon)
                  \left\{ \hat{e}_z
                          \left| \phi_{\nu}(\vec r_{\perp}) \right|^2
                          \mbox{Im}
                          \left[ \left( \psi^{(s)}(\epsilon,z) \right)^*
                                 \frac{\partial \psi^{(s)}(\epsilon,z)}
                                      {\partial z}
                          \right]
                  \right.
\nonumber \\
             & & \hspace*{3cm}
                +\left.\left| \psi^{(s)}(\epsilon,z) \right|^2
                         \mbox{Im}
                          \left[ \phi^*_{\nu}(\vec r_{\perp})
                                 \nabla_{\vec r_{\perp}}
                                 \phi_{\nu}(\vec r_{\perp})
                          \right]
                  \right\}
                  \bar{n}_{\nu s}(\epsilon),
\label{jj-2}
\end{eqnarray}
respectively, where $\bar{n}_{\nu s}(\epsilon)$ is 
the mean occupation number of the single
particle state $(\nu,s,\epsilon)$,
\begin{equation}
\bar{n}_{\nu s}(\epsilon) = \mbox{Tr} \left[ \hat \rho \,
                                             c^\dagger_{\nu s}(\epsilon)
                                             c_{\nu s}(\epsilon)
                                      \right] 
\end{equation}
and $\hat{e}_z$ is the unity vector of the $z$ axis.

Further we analyze the component of the current density 
on the transport direction. 
Replacing in Eq. (\ref{jj-2}) the scattering function by their expressions 
(\ref{psi-out-1}) and (\ref{psi-out-2}),
and $g_s(\epsilon)$ by Eq. (\ref{gs}) we find
\begin{eqnarray}
j_z(\vec r_{\perp},z)
           &=& \frac{2 e}{ h}
               \sum_{\nu}
               \left| \phi_{\nu}(\vec r_{\perp}) \right|^2
               \int_{V_M}^{\infty} d \epsilon \; 
               \left\{ \bar{n}_{\nu 1}(\epsilon)
                       \left[ 1 - \left| \tilde{\bf S}_{11}(\epsilon)\right|^2 
                       \right]
                      -\bar{n}_{\nu 2}(\epsilon)
                       \left| \tilde{\bf S}_{21}(\epsilon)\right|^2
               \right\}
\end{eqnarray}
in the source contact region ($z < -d$) and
\begin{eqnarray}
j_z(\vec r_{\perp},z)
           &=& \frac{2 e}{h}
               \sum_{\nu}
               \left| \phi_{\nu}(\vec r_{\perp}) \right|^2
               \int_{V_M}^{\infty} d \epsilon \; 
               \left\{ \bar{n}_{\nu 1}(\epsilon)
                       \left| \tilde{\bf S}_{12}(\epsilon)\right|^2  
                      -\bar{n}_{\nu 2}(\epsilon)
                       \left[ 1 - \left| \tilde{\bf S}_{22}(\epsilon)\right|^2
                       \right]
               \right\} 
\end{eqnarray}
in the drain contact region ($z>d$).
The nondegenerate energy levels between $V_m$ and $V_M$
do not contribute to the current because on the one side of the
scattering region the corresponding
scattering functions decay exponentially 
and on the other side the reflection coefficient is 1. 
For $\epsilon > V_M$ the current scattering 
matrix $\tilde{\bf S}(\epsilon)$ is an unitary matrix 
and in turn the component
of the current density in the transport direction 
can be written as
\begin{eqnarray}
j_z(\vec r_{\perp},z)&=& \frac{2 e}{h}
                              \sum_{\nu}
                              \left| \phi_{\nu}(\vec r_{\perp}) \right|^2
                              \int_{V_M}^{\infty} d \epsilon \; 
                               T(\epsilon)
                               \left[ \bar{n}_{\nu 1}(\epsilon)
                                     -\bar{n}_{\nu 2}(\epsilon)
                               \right],
\label{jj-}
\quad |z| > d
\end{eqnarray}
where $T(\epsilon)$ is the transmission probability.
The current density is actually independent on $z$ in the contact regions
and has the same value in the source and drain contacts proving the current 
conservation through the system.

To calculate the mean value of the occupation number
$\bar{n}_{\nu s}(\epsilon)$ of the single particle state
$(\nu,s,\epsilon)$
we need some considerations about the conduction electrons in
nanostructures. The electron gas analyzed above 
extends in fact only in a domain whose dimensions are
comparable with the phase coherence length of the nanostructure 
and are larger than the size of the scattering region.
The rest of the heterostructure is usually
a highly doped semiconductor or a metal and acts as a particle and energy
reservoir for the considered electron gas. 
If the phase coherence length is large enough 
so that the electron gas can
be taken as a macroscopic system, 
the particle number and the energy of this system
are practically constant in the limit of thermodynamic equilibrium.
That means that the electron gas is quasi-isolated
and the above expressions for the charge
and current densities remain valid. The advantage of considering the
system of
electrons in contact with a particle- and energy bath is that
the mean value of the occupation number
is given by the Fermi-Dirac distribution function \cite{huang},
\begin{equation}
\bar{n}_{\nu s}(\epsilon) = f_{FD}(\epsilon+E_\perp^\nu-\mu),
\label{FD}
\end{equation}
where $\mu$ is the chemical potential of the system at thermodynamic
equilibrium, fixed by the doping in the contact regions of the
heterostructure.
The expression (\ref{FD}) of the mean value of the occupation number
assures that the current is zero at thermodynamic equilibrium.

\subsection{LANDAUER-B\"UTTIKER FORMALISM}

To calculate the transport properties in the
Landauer-B\"uttiker formalism \cite{lanbue},
electrons can be thought of as two noninteracting Fermi-gases:
First, the electrons coming from the source contact which occupy
the single-particle scattering states
with $s=1$ according to
the  Fermi-Dirac distribution function $f_{FD} (E-\mu_1)$,
\begin{equation}
\bar{n}_{\nu 1}(\epsilon) = f_{FD}(\epsilon+E_\perp^\nu-\mu_1), 
\label{FD1}
\end{equation}
where $\mu_1$ is
the  chemical potential of the source contact. Second,
the electrons coming from the drain contact which occupy the single particle
states indexed by $s=2$ according to
the  Fermi-Dirac distribution function $f_{FD} (E-\mu_2)$,
\begin{equation}
\bar{n}_{\nu 2}(\epsilon) = f_{FD}(\epsilon+E_\perp^\nu-\mu_2), 
\label{FD2}
\end{equation}
$\mu_2= \mu_1-eU_{sd}$
being the chemical potential of the drain contact.
The potential difference $\mu_1-\mu_2$
results from an externally applied drain-source voltage, $U_{sd}$.

Using these basic assumptions of the Landauer-B\"uttiker formalism
and the normalization condition (\ref{basis-2D}) of the functions
$\phi_\nu(\vec r_\perp)$
we determine from Eq. (\ref{qq-2}) the electronic charge 
$Q=\int d \vec r \, q(\vec r)$ inside the 
scattering system, 
\begin{equation}
Q =-\frac{2em^*}{\hbar^2} 
    \sum_s \int_{V_s}^{\infty} d \epsilon\; 
      \frac{P_s(\epsilon)}{k_s(\epsilon)}
      \sum_\nu f_{FD}(\epsilon+E_\perp^\nu-\mu_s),
\label{charge}
\end{equation}
where 
\begin{equation}
P_s(\epsilon) 
=\int_{-d}^d dz \left|\psi^{(s)}(\epsilon,z)\right|^2
\label{Peps}
\end{equation}
is the particle probability distribution.
For a free electron system in a plane perpendicular to the transport
direction index $\nu$ should be replaced by a 2D wave vector 
$\vec k_\perp$ and the electron charge in the scattering region becomes
\begin{equation}
Q =-A \frac{e}{\pi} \left( \frac{m^*}{\hbar^2} \right)^2
    \sum_s \int_{V_s}^{\infty} d \epsilon\; 
      \frac{P_s(\epsilon)}{k_s(\epsilon)}
      \int dE \, f_{FD}(E-\mu_s),
\label{charge2DEG}
\end{equation}
where $E$ denotes the total energy of the electron
and $A$ is the lateral area of the system.
It is the great advantage of using the R matrix formalism to enable 
the analytical calculation of $P_s(\epsilon)$
inside the scattering system, as shown  in Appendix \ref{A-ortho}.
Its expression is given by Eq. (\ref{rho-mat-ee})
and using the expressions (\ref{Psi-mat}-\ref{Psi-S-mat})
for $\Psiv$ and $\Psiv_S$ we obtain
\begin{equation}
P_s(\epsilon) 
= \frac{1}{2\pi}
  \mbox{Im} \left[ \frac{1}{g_s(\epsilon)}
                   \left( \tilde{\bf S}^\dagger(\epsilon)
                          \frac{d \tilde{\bf S}}{d \epsilon}
                   \right)_{ss}
                  +\frac{\tilde{\bf S}_{ss}(\epsilon)}{k_s(\epsilon)}
            \right]
\end{equation}
for $\epsilon > V_M$,
\begin{equation}
P_1(\epsilon)
 = \frac{1}{2 \pi}
   \mbox{Im}\left[ \frac{1}{g_1(\epsilon)} \tilde{S}_{11}^*(\epsilon)
                   \frac{d \tilde{S}_{11}}{d \epsilon}
                  +\frac{\tilde{S}_{11}(\epsilon)}{k_1(\epsilon)}
             \right]
  -\frac{1}{4 \pi}
   \frac{k_1(\epsilon)}{|k_2(\epsilon)|^2}
   \left| \tilde{\bm S}_{21}(\epsilon) \right|^2
\end{equation}
for $\epsilon < V_M$ in the case $V_1 < V_2$, and
\begin{equation}
P_2(\epsilon)
= \frac{1}{2 \pi}
  \mbox{Im}\left[ \frac{1}{g_2(\epsilon)} \tilde{S}_{22}^*(\epsilon)
                  \frac{d \tilde{S}_{22}}{d \epsilon}
                 +\frac{\tilde{S}_{22}(\epsilon)}{k_2(\epsilon)}
            \right]
 -\frac{1}{4\pi}
  \frac{k_2(\epsilon)}{|k_1(\epsilon)|^2}
  \left| \tilde{\bm S}_{12}(\epsilon) \right|^2
\end{equation}
for $\epsilon < V_M$ in the case $V_2 < V_1$.
The above relations demonstrate that 
the electronic charge accumulated inside the scattering system is
only  given by the elements of the scattering matrix.

To describe the transport properties of nanostructures we also need to
analyze the current through the system,
$I = \int d \vec r_{\perp} j_z(\vec r_{\perp},z)$
as a functions of the applied drain-source voltage.
Using Eqs. (\ref{FD1}-\ref{FD2})
we obtain from Eq. (\ref{jj-})
the current  as
\begin{equation}
I = \frac{2 e}{h}
    \int_{V_M}^{\infty} d \epsilon \; T(\epsilon)
    \sum_{\nu} \left[ f_{FD}(\epsilon+E_\perp^\nu-\mu_1)
                     -f_{FD}(\epsilon+E_\perp^\nu-\mu_2)
               \right]
\label{current}
\end{equation}
for $|z|>d$, where $T(\epsilon)$ is the transmission probability
characterizing the scattering region of the system.
This is the current between the source and drain contacts. 
We can also define
a current in the lateral directions as results from Eq. (\ref{jj-2}),
but this current is zero for the systems considered in this paper.
For the heterostructures with total lateral confinement,
the energy levels $E_{\perp}^{\nu}$
are nondegenerate and the time translational invariance
allows us to choose the function $\phi_{\nu}(\vec r_{\perp})$ as real.
It results immediately that
$\mbox{Im} \left[ \phi^*_{\nu}(\vec r_{\perp})
           \nabla_{\vec r_{\perp}}
           \phi_{\nu}(\vec r_{\perp}) \right] =0$ and the
lateral component of the current density vanishes.
In the opposite limit,
for a free electron gas in a plane perpendicular to the
transport direction
we can define a 2D wave vector $\vec k_{\perp}$ and
$\phi_{\nu}(\vec r_{\perp}) \rightarrow \phi(\vec k_\perp,\vec r_\perp)
= e^{i \vec k_{\perp} \vec r_{\perp}}/2 \pi$. 
It follows in a straightforward manner that
$\mbox{Im} \left[ \phi^*_{\nu}(\vec r_{\perp})
           \nabla_{\vec r_{\perp}}
           \phi_{\nu}(\vec r_{\perp}) \right] 
\rightarrow \vec k_{\perp}/(2 \pi)^2 $
and the lateral component of the current density becomes zero.

For determining the conductance 
it is convenient to write the expression (\ref{current}) of
the current into an equivalent form
\begin{equation}
I =\frac{2 e}{h}
        \int_{V_M}^{\infty} d E \;
              \left[ f_{FD}(E -\mu_1)
                    -f_{FD}(E -\mu_2)
              \right]
        \sum_{\nu} T(E - E_\perp^\nu)
                   \theta(E - E_\perp^\nu),
\label{c-dot}
\end{equation}
where the integration is made over the total
energy of the electron.
This expression has the advantage that it directly yields the conductance.
Each energy level $E_{\perp}^{\nu}$
defines a channel for the electron transport \cite{bue85}
and it is usually said that the channels which satisfy the condition
$0<E_\perp^\nu<E_F$ are open channels because they contribute to the
current.
The $\theta$-function in Eq. (\ref{c-dot})
serves to remove the channels with exponentially
decaying wave functions in the contacts, the so called closed channels for
transport. 
 In the linear response regime
($U_{sd} \rightarrow 0$,  $V_1 = V_2 \equiv 0$)
 and for low temperatures ($T\rightarrow 0$),
we can expand the Fermi-Dirac function $f_{FD}(E -\mu_2)$
in a Taylor series around $E-\mu_1$,
and thus obtain \cite{prb01} from Eq.\ (\ref{c-dot})
\begin{equation}
G = \frac{2 e^2}{h}
    \sum_{\nu} T(E_F-E_{\perp}^{\nu})
               \theta(E_F-E_{\perp}^{\nu}),
\label{conductio-gen}
\end{equation}
where $T(\epsilon)= \left|\tilde {\bf S}_{12}(\epsilon)\right|^2$
is the transmission probability 
determined solely by the one-dimensional scattering problem
Eq.\ (\ref{eq-S-1D})
and $E_F$ is the Fermi energy of the electron gas in the
source contact.

\section{RESONANCES IN TRANSPORT}
\label{res}

The notion of resonances, representing long-lived intermediate 
states of an open system to which bound states of its closed 
counterpart are converted due to coupling to continuum,
is one of the most fundamental concepts in the domain of
quantum scattering \cite{kukulin}. On a formal level 
resonances show up as poles of the scattering matrix occurring at 
complex energies 
$\bar{\epsilon}_{0} = \epsilon_{0}-i \Gamma/2$,
where $\epsilon_{0}$ and $\Gamma$ are called position 
and width of the resonance, respectively \cite{fyodorov}.

The causality condition leads to analytic properties of the 
scattering matrix
when the energy or the wave vector of the electron are extended to
complex values \cite{bohm}.
To calculate the poles of the scattering matrix we exploit these
properties which are briefly summarized in the following.
$\tilde{S}(\epsilon)$ is a meromorphic function on the 
two-sheeted Riemann surfaces - one for positive values of the imaginary
part of the wave vector and the other for negative values -
with a branch point at $\epsilon=0$
and a cut from 0 to $\infty$. 
Bound states poles lie on the negative real axis of the 
"physical sheet" and, except for them,
$\tilde{S}(\epsilon)$ is an analytical function on the physical 
sheet\cite{kampen}. The resonance poles
lie on the second, "unphysical sheet" and come from the possible zeros 
on the first sheet \cite{bohm}.
As usually done, we assume these poles to be simple
\cite{noeckel}.
\begin{figure}[h]
\begin{center}
\noindent\includegraphics[width=4.05in]{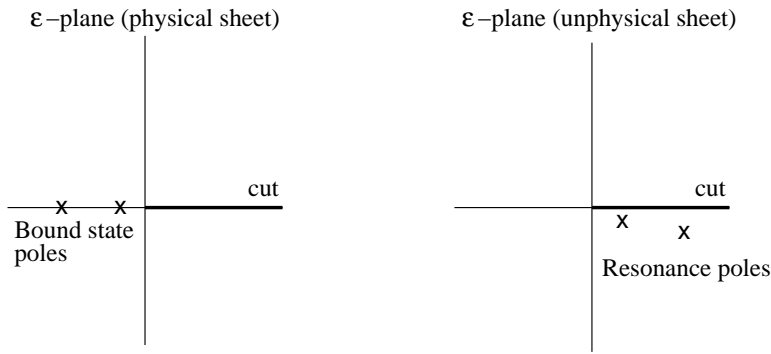}
\end{center}
\caption{Left side: "Physical sheet" of the complex energy plane
corresponding to the values of the wave vector 
$k=\sqrt{2 m^* \epsilon/\hbar^2}$
with $\mbox{Im}[k] > 0$. 
Poles of the scattering matrix associated with the bound states.
Right side: "Unphysical sheet" of the complex energy plane
corresponding to the values of the wave vector with $\mbox{Im}[k] <0$. 
Poles of $\tilde{S}(\epsilon)$ associated to resonances.
}
\label{E_plane}
\end{figure}

Our development of the resonance theory starts with the  
{\bf R} matrix representation of the {\bf S} matrix
which 
is particularly well suited to the description 
of narrow resonances \cite{kukulin}.
As demonstrated in Sec. \ref{smat},
the two matrices are related to each other through 
Eq. (\ref{rel-StR}).
It is obvious that the resonance poles 
correspond to the zeros of the denominator function
in the expression (\ref{rel-StR})  of the $\tilde{\bf S}$ matrix,  
\begin{equation}
\det[{\bf 1}+i\Omegav(\epsilon_0-i \Gamma/2)]
= 0.
\label{poles}
\end{equation}

According to the definition (\ref{Omega}) 
the $\Omegav$ matrix elements have single poles 
for each $\epsilon=\epsilon_l$,
$l \ge 1$, i.e. for each eigenenergy of the Hamiltonian corresponding
to the scattering system isolated from the contacts.
The Wigner-Eisenbud energies $\epsilon_l$ are real
and are associated to the bound states.
If the system is coupled to contacts
the bound states are transformed into scattering states
(at most with a few exceptions depending on the particularities of $V(z)$)
and the poles should migrate in the lower part of the "unphysical sheet"
of the complex energy plan. 
Further we demonstrate that the Wigner-Eisenbud energies, $\epsilon_l$,
are not solutions of 
Eq. (\ref{poles}) and therefore they are not singularities  
of the $\tilde{\bf S}$ matrix.
The matrix $\Omegav$ 
is split into a resonant part which contains 
$(\epsilon-\epsilon_\lambda)^{-1}$ and a regular matrix 
$\Omegav_\lambda$,
\begin{equation}
\Omegav(\epsilon) = \frac{\omegav_\lambda(\epsilon)}
                              {\epsilon-\epsilon_{\lambda}}
                        +\sum_{\mbox{\footnotesize $\begin{array}{c}
                                                     l=1 \\
                                                     l \ne \lambda
                                                    \end{array}$}}
                             ^{\infty}
                         \frac{\omegav_l}{\epsilon-\epsilon_l}
                       = \frac{\omegav_\lambda(\epsilon)}
                              {\epsilon-\epsilon_{\lambda}}
                        +\Omegav_\lambda(\epsilon).
\label{omel}
\end{equation}
This decomposition and the property of $\omegav_\lambda$
to have the determinant zero lead to
\begin{equation}
\det[{\bf 1} + i \Omegav]
=\frac{ \epsilon-\epsilon_\lambda
       -\bar{{\cal E}}_{\lambda}(\epsilon)}
      {\epsilon-\epsilon_\lambda}
 {\cal D}_{\lambda}(\epsilon)
\end{equation}
with
$
\bar{\cal E}_{\lambda}(\epsilon)
   = -i \mbox{Tr}[\omegav_{\lambda}
                         ({\bf 1}+i \Omegav_{\lambda})^{-1}]
$
and
$
{\cal D}_{\lambda}(\epsilon)
         = \mbox{det} \left[{\bf 1} + i \Omegav_{\lambda}
                      \right].
$
Now we can express the ${\bf \tilde{S}}$-matrix 
in terms of $\Omegav_{\lambda}$ and $\omegav_{\lambda}$,
\begin{equation}
\tilde{\bf S}(\epsilon) = \frac{{\bf Z}_{\lambda}(\epsilon)}
                               { \epsilon-\epsilon_{\lambda}
                                -\bar{{\cal E}}_{\lambda}(\epsilon)},
\label{S-matrixl}
\end{equation}
where the matrix ${\bf Z}_{\lambda}(\epsilon)$ is defined as
$
{\bf Z}_{\lambda}(\epsilon) = (\epsilon-\epsilon_\lambda)
                              \left[-1 - \det[\Omegav]
                                    +i \left( \Omegav -\Omegav^{-}
                                       \right)
                              \right]/{\cal D}_{\lambda}(\epsilon)
$
with $\Omegav^{-} = \Omegav^{-1} \det[\Omegav]$.
In principle, the solutions of the equation 
 ${\cal D}_{\lambda}(\epsilon) = 0$
can not be associated with resonance poles of the $\tilde{\bf S}$ matrix.
As follows from its  definition 
 $\bar{\cal E}_{\lambda}(\bar{\epsilon})$
becomes infinite for ${\cal D}_\lambda(\bar{\epsilon}) \rightarrow 0$.
We cannot argue that ${\cal D}_\lambda$ and
 $\det[{\bf 1} + i \Omegav]$  vanish
at the same points
and it is convenient to include ${\cal D}_\lambda(\epsilon)$
in ${\bf Z}_{\lambda}(\epsilon)$. 

The matrix ${\bf Z}_{\lambda}$ and the function $\bar{\cal E}_{\lambda}$
are related to each other through the unitarity requirement
for the $\tilde{\bf S}$ matrix which gives
\begin{equation}
  {\bf Z}_{\lambda} {\bf Z}_{\lambda}^{\dagger}
= {\bf Z}_{\lambda}^{\dagger} {\bf Z}_{\lambda}
= \left| \epsilon - \epsilon_{\lambda} 
        - \bar{\cal E}_{\lambda}(\epsilon) 
  \right|^2,
\label{Zfunk}
\end{equation}
for $\epsilon > V_M$.

The representation of the {\bf S} matrix in
Eq.\ (\ref{S-matrixl}) is an exact reformulation
of Eq.\ (\ref{rel-StR}) and has the advantage of directly yielding
the equation
\begin{equation}
  \bar{\epsilon}_0 - \epsilon_{\lambda}
- \bar{\cal E}_{\lambda}(\bar{\epsilon}_0) = 0
\label{pole1}
\end{equation}
to determine the positions
$\bar{\epsilon}_0 = \epsilon_0 - i \Gamma/2$
of the poles in the complex energy plane. 
It is obvious that Eq. (\ref{S-matrixl})
has no singularities for real energies in the interval
$(\epsilon_{\lambda-1},\; \epsilon_{\lambda+1})$.
The coupling to  the contacts leads to a nonzero
imaginary part of the resonance energy and
the stronger the coupling the larger the difference
between the Wigner-Eisenbud energy $\epsilon_\lambda$
and the resonant energy $\bar{\epsilon}_{0\lambda}$, $\lambda \ge 1$.
Each Wigner-Eisenbud energy was associated a resonance energy.

In the following we analyze the narrow
transport resonances for which $\Gamma$ is a small quantity
and the line shape is given by an asymmetric Fano function.
Therefore, as a basic assumption
for our theory of the resonant Fano line shape, we require the validity
of  the linearization of
$\bar{{\cal E}}_{\lambda}$ and implicitly ${\bf Z}_{\lambda}$
in a domain of the
complex energy plane that includes the pole $\bar{\epsilon}_0$
and the part of the real axis which contains the
transmission peak associated with the resonance, i.e the resonance domain.
To obtain the line shape of the resonance we employ a formal expansion
of the $\tilde{\bf S}$-matrix as given in Eq.\ (\ref{S-matrixl}) in a 
Laurent series around the pole $\bar{\epsilon}_0 = \epsilon_0 - i \Gamma/2$
and neglect the derivatives 
up to the second order of $\bar{{\cal E}}_{\lambda}$ and
${\bf Z}_{\lambda}$ 
at the points $\bar{\epsilon}_0$ and $\epsilon_0$ \cite{prb01,thesis}.
Thus, the $\tilde{\bf S}$ matrix has the form
\begin{equation}
\tilde{\bf S}(\bar{\epsilon}) \simeq \frac{1}
                                     { \bar{\epsilon} -\epsilon_0 
                                      +i \Gamma/2}
                                \frac{ {\bf Z}_{\lambda}(\epsilon_0)
                              -i \Gamma/2
                               \left. \frac{d {\bf Z}_{\lambda}}
                                           {d \epsilon}
                               \right|_{\epsilon=\epsilon_0}}
                             {1 - \left. \frac{d \bar{\cal E}_{\lambda}}
                                              {d \epsilon}
                                  \right|_{\epsilon=\epsilon_0}}
                            +\frac{\left. \frac{d {\bf Z}_{\lambda}}
                                          {d \epsilon}
                              \right|_{\epsilon=\epsilon_0}}
                             {1 - \left. \frac{d \bar{\cal E}_{\lambda}}
                                              {d \epsilon}
                                  \right|_{\epsilon=\epsilon_0}}.
\end{equation}
Eq.\ (\ref{S-matrixl})
ensures that the ${\bf S}$ matrix is an analytic function
in the resonance domain excepting the pole,
condition 
which is required for the existence of the Laurent series.
After the linearization of $\bar{\cal E}_{\lambda}$
in Eq. (\ref{pole1}) we find 
$1 - \left. d \bar{\cal E}/d\epsilon \right|_{\epsilon=\epsilon_0}
=\left( \epsilon_0 - \epsilon_{\lambda} 
       -\bar{\cal E}_{\lambda}(\epsilon_0) \right) / i \Gamma/2$
and then the expression of the $\tilde{\bf S}$ matrix inside the
resonance domain becomes
\begin{equation}
{\bf \tilde{S}}(\epsilon) \simeq i \frac{ \tilde{\bf S}(\epsilon_0)
                                         -\tilde{\bf S}_{bg}}
                                        {e+i}
                                 +\tilde{\bf S}_{bg},
\label{laurent}
\end{equation}
where $e = 2(\epsilon-\epsilon_0)/\Gamma$
and
\begin{equation}
\tilde{\bf S}_{bg} = \frac{i \Gamma/2}
                          { \epsilon_0-\epsilon_{\lambda}
                           -\bar{\cal E}_{\lambda}(\epsilon_0)}
                     \left. \frac{d {\bf Z}_{\lambda}} {d \epsilon}
                     \right|_{\epsilon= \epsilon_0}.
\label{Sbg}
\end{equation}
Eq.\ (\ref{laurent}) has 
a standard form 
\cite{bohm,noeckel,bue88,landau},
but we can provide here an explicit expression  
for the nonresonant component of the scattering matrix,
$\tilde{\bf S}_{bg}$. 
The first term in Eq. \ (\ref{laurent})
represents the resonant part of $\tilde{\bf S}$.
For each element of the matrix $\tilde{\bf S}$
it is seen from Eq.\ (\ref{laurent}) that
the resonant part undergoes a phase change of $\pi$ when the energy
passes the resonance.  In general, this produces a change between constructive
and destructive superposition of the resonant and the nonresonant part.
Therefore, an asymmetric line is obtained.

The  resulting
expression (\ref{laurent}) 
preserves the unitarity of the scattering matrix only in linear 
order $e$ \cite{prb01}: 
\begin{equation}
\tilde{\bf S} \tilde{\bf S}^{\dagger}
=\tilde{\bf S}^{\dagger} \tilde{\bf S}
\simeq {\bf 1} + (\deltav-{\bf 1})
                 \frac{e^2}{e^2+1}.
\label{unitar}
\end{equation}
We expect that our approximation is valid as
long as the second term on the right hand side of
Eq.\ (\ref{unitar}) is small compared to 1, i.e. 
the deviation of $\tilde{\bf S}$ from unitarity is small.
This way, for each maximum we can estimate the range of validity
for our approximation through the requirement
$(\deltav-{\bf 1})_{ij} \, e^2 /(e^2+1) \ll 1$, $i,j=1,2$.

\section{APPLICATIONS}

\subsection{CONDUCTANCE THROUGH A QUANTUM DOT} 
\label{conductance}

\begin{figure}[h]
\begin{center}
\noindent\includegraphics[width=5.5in]{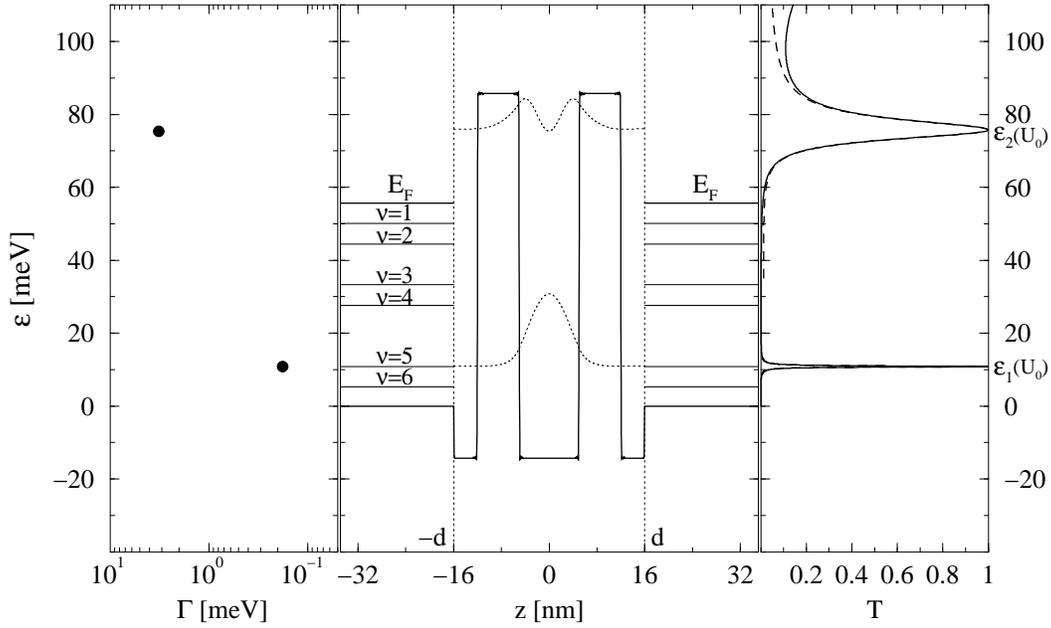}
\end{center}
\caption{Middle: Total potential energy
of a double-barrier structure, $V(z)-eU_g$, for a value of the gate voltage
which corresponds to a maximum in conductance, $U_g=U_0=14.28mV$. 
The potential steps
of height $-eU_0$ at $|z|= d$ ($d=16 nm$) result
from the voltage applied to the plunger gate. In dotted lines
the wave functions $|\psi^{(1)}|^2$ at the resonant energies
$\epsilon_i(U_0)$, $i =1,2$. 
The energies $E_F-E_\perp^\nu$ (solid lines) are drawn 
in the contact regions, 
where $E_\perp^\nu$, $\nu=1,6$ correspond
to an infinite 2D quantum well with the dimensions $16nm \times 16nm$.
Left: Positions of the resonance poles of the $\tilde{\bf S}$ matrix
in the complex energy plane.
Right: Transmission $T(U_0;\epsilon)$ vs. energy
calculated for the scattering potential energy $V(z)-eU_0$:
numerical calculation (solid lines) and Fano approximation (dashed lines).
}
\label{cond-fig1}
\end{figure}

As a first application of our resonance theory of transport 
we calculate the conductance through a quantum dot embedded in a quantum
wire\cite{prb01}. 
The potential energy experienced by electrons has two components:
the $z$-independent lateral confinement potential energy
$V_{\bot}(\vec r_{\bot})$, which provides the one-dimensional character
of the structure, and
$V(z)$ a double barrier potential separating the 
quantum dot
from the rest of the system (Fig. \ref{cond-fig1}).

In the linear response regime
($V_{sd} \rightarrow 0$,  $V_1 = V_2 \equiv 0$)
and for low temperatures ($T\rightarrow 0$),
the conductance through the dot 
is given by Eq. (\ref{conductio-gen}) as a
superposition of transmission curves 
$T(E_F-E_{\perp}^{\nu})$, where $\nu$ corresponds to an open channel for
the transport. 

In experiments $G$ and therefore $T$ is probed at different energies
by varying the voltage of an  additional plunger gate.
In case of lateral tunneling, this additional
gate is a top gate \cite{goeres,lat} and, in the case of
vertical tunneling, it is a side gate \cite{tarucha95}.
We use the following idealization
for the total potential in presence of a varying  gate voltage:
The external potential created by the charges at the gate
is screened out completely in the heavily doped contacts
($|z|>d$) so that the total potential and
$E_F$ remain unchanged. In the scattering area
($|z|<d$) the total potential
can be idealized for small variations of the gate voltage as a
 varying potential energy offset $eU_g$, so that 
$V(z) \rightarrow V(z) - eU_g$.
As shown in Fig.\ \ref{cond-fig1}
the transmission probability of the double barrier system
depends on the gate potential $U_g$
and therefore the conductance $G$ varies with $U_g$.

As illustrated in Fig.\ \ref{cond-fig1} for small energies
($\epsilon < \max[V(z)]$) 
the transmission is generally small and may have some isolated peaks at
$\epsilon_i$. The peak position depends on the total potential energy
experienced by electrons and therefore $\epsilon_i=\epsilon_i(U_g)$.
The line shape of a transmission peak is deduced from 
Eq.\ (\ref{laurent}) and is given by
\begin{equation}
T(\epsilon) = \left| (\tilde{\bf S}(\epsilon))_{12} \right|^2
            \simeq T_{bg} \frac{[e+\mbox{Re}(q)]^2 + [\mbox{Im}(q)]^2}
                          {e^2 + 1},
\label{lines}
\end{equation}
where $T_{bg} = \left|(\tilde{\bf S}_{bg})_{12}\right|^2$
is the background transmission.
The right hand side of Eq.\ (\ref{lines}) is a Fano distribution 
with a complex asymmetry parameter\cite{prb01}
\begin{equation}
q  =  i \left( \tilde{\bf S}(\epsilon_0) \right)_{12}
        \left( \tilde{\bf S}_{bg} \right)_{12}^{-1}.
\label{asym}
\end{equation}
As shown in Fig. \ref{cond-fig1} the Fano approximation
for the line shape of the transmission
gives a very good description 
even for the asymmetric peaks as long as the peaks are quasi-isolated.
The background matrix $\tilde{\bf S}_{bg}$ in Eq. (\ref{laurent}) 
is often assumed to be absent, leading to a symmetrical Wigner-Breit line shape
of the transmission\cite{WB},
but it is obviously that such a description fails  
in the case of an asymmetric peak as the second peak in 
Fig.  \ref{cond-fig1}.

\begin{figure}[h]
\begin{center}
\noindent\includegraphics[width=4.0in]{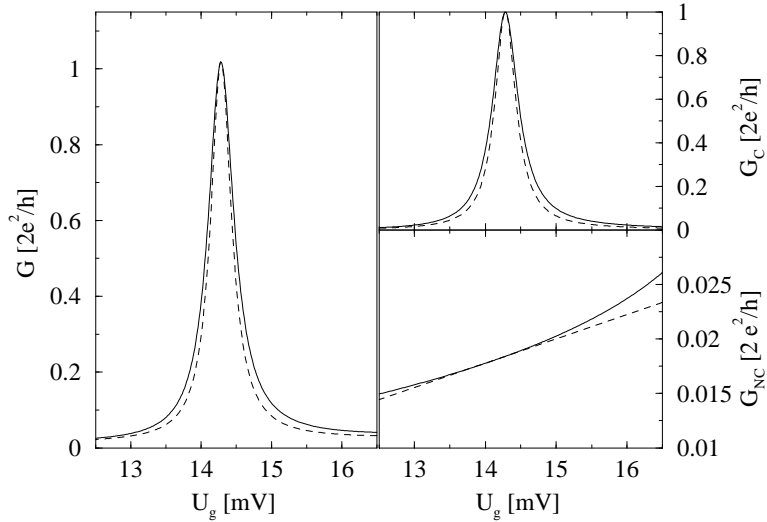}
\end{center}
\caption{Left: Conductance vs. gate voltage around the maximum at
$U_0=14.28\;mV$: complete calculation Eq. (\ref{conductio-gen}) (solid line) 
and approximative value $G=G_{C}+G_{NC}$
given by Eqs. (\ref{GG-C-3}) and (\ref{GG-NC-3}) (dashed line).
Right upper part: Coherent contributions to the conductance around
maximum: complete calculation Eq. (\ref{GG-C}) (solid line)
and Fano approximation Eq. (\ref{GG-C-3}) (dashed line).
Right lower part: Noncoherent contributions to the conductance around
maximum: complete calculation Eq. (\ref{GG-NC}) (solid line)
and linear approximation Eq. (\ref{GG-NC-3}) (dashed line). 
}
\label{cond-fig2}
\end{figure}

If the Fermi energy is smaller than the maximum of the barriers 
and if the  external voltages $U_g$ are quite small,
only the isolated peaks participate to the transport. According to 
Eq. (\ref{conductio-gen}) the conductance has a maximum each time 
the transmission of an open channel has a peak,
\begin{equation}
\epsilon_{i_0}(U_0) = E_F-E_{\perp}^{\nu_0}.
\label{res-cond}
\end{equation}
Thus, to each conductance maximum at $U_g=U_0$ 
a pair index $(\nu_0,i_0)$ is assigned,
where $\nu_0$ is the channel index and $i_0$ the number of the
maximum in the curve $T(U_0;\epsilon)$.
As shown in Fig. \ref{cond-fig1} each index $i_0$ can be associated with a
pole of the scattering matrix and we can conclude that a maximum in
conductance is characterized by a resonant channel $\nu_0$ and a resonant
pole of the $\tilde{\bf S}$ matrix, $\bar{\epsilon}_{i_0}$.
On this basis one can split the contribution to the conductance into a
coherent resonant part,
\begin{equation}
G_{C}(U_g) = \frac{2 e^2}{h}
                  T(U_g;E_F-E_{\perp}^{\nu_0}),
\label{GG-C}
\end{equation}
and a slowly varying noncoherent part 
\begin{equation}
G_{NC}(U_g) = \frac{2 e^2}{h} \sum_{\nu \neq \nu_0}
                T(U_g;E_F-E_{\perp}^{\nu}),
\label{GG-NC}
\end{equation}
in which the absolute squares of the transmission
coefficients are added without phase-information.
In the above expression only the contributions of the open channels 
are counted. 
Fig. \ref{cond-fig2} illustrates 
the conductance through the dot and its coherent and noncoherent
contributions around a maximum at $U_g=U_0$.

In case of a narrow conductance peak the R-matrix representation of the
scattering matrix allows for obtaining
an explicit dependence $G=G(U_g)$ in the vicinity of the maximum,
$U_g=U_0+\delta U$. 
By the transformation $V(z) \rightarrow V(z)-eU_g$
the Wigner-Eisenbud energies $\epsilon_l$ [Eq. (\ref{WE-eq})]
become $\epsilon_l-eU_g$
but the Wigner-Eisenbud functions $\chi_l$ 
remain unchanged. Then, according to Eq. (\ref{Omega}) the matrix
$\Omegav(U_g;\epsilon)$ can be approached by 
$\Omegav(U_0;\epsilon+e \delta U)$. 
Thus, from Eq. (\ref{rel-StR}) follows that 
$T(U_g;\epsilon) \simeq T(U_0;\epsilon+e \delta U)$.
Inserting condition (\ref{res-cond}) in the above relation
we obtain the coherent and noncoherent contributions to the
conductance around the maximum, 
\begin{equation}
G_{C}(U_g) \simeq \frac{2 e^2}{h}
             T(U_0;\epsilon_{i_0} + e \delta U),
\label{GG-C-2}
\end{equation}
and 
\begin{equation}
G_{NC}(U_g) \simeq \frac{2 e^2}{h} \sum_{\nu \neq \nu_0}
              T(U_0;\epsilon_{i_0}+E_\perp^{\nu_0}-E_\perp^\nu+e \delta U), 
\label{GG-NC-2} 
\end{equation} 
respectively,
where $\epsilon_{i_0}$ is the position of the $i_0$-th transmission maximum
for the potential energy $V(z)-eU_0$.

Using Eq. (\ref{lines})
the coherent contribution to the conductance in the vicinity of the resonance
is obtained
as a Fano function with the complex asymmetry parameter $q$ defined by
(\ref{asym}),
\begin{equation}
G_C(U_g) \simeq G_{bg}
          \frac{ \left[ \tilde{e} + \mbox{Re}(q) \right]^2
                +\left[ \mbox{Im}(q) \right]^2}
               {\tilde{e}^2 + 1}.
\label{GG-C-3}
\end{equation}
Here
$\tilde{e} = 2e(U_g-U_0)/\Gamma_{i_0}$
is a function of the plunger gate potential $U_g$, the
resonance position $U_0$ and the resonance width $\Gamma_{i_0}$
calculated as the imaginary part of the $i_0$-th pole
of the scattering matrix for the potential $V(z)-eU_0$.
The background coherent contribution is related to the background
transmission through
$ G_{bg} = (2 e^2/h) T_{bg}$.

As shown in Fig. \ref{cond-fig2}
the noncoherent part of the conductance
varies slowly inside the resonance domain and we can expand this function
in a Taylor series for $U_g \sim U_0$. One obtains
\begin{equation}
G_{NC}(U_g) \simeq \frac{2 e^2}{h} \sum_{\nu \neq \nu_0}
              T(U_0;\epsilon_{i_0}+E_\perp^{\nu_0}-E_\perp^\nu)
             +e \delta U \frac{2 e^2}{h} \sum_{\nu \neq \nu_0}
              \left. \frac{d T}{d \epsilon} 
              \right|_{\epsilon=\epsilon_{i_0}+E_\perp^{\nu_0}-E_\perp^\nu}.
\label{GG-NC-3}
\end{equation}
As illustrated in Fig. \ref{cond-fig2}, Eqs. (\ref{GG-C-3}) and
(\ref{GG-NC-3}) provide a very good description of the two 
contributions to the conductance of the dot.
If the overlap of the peaks is very small,
the noncoherent  contribution to the conductance
can be considered even as constant \cite{prb01}.

We can conclude that the most general form of a conductance profile is a Fano
line with a complex asymmetry parameter superposed on a linear function
which usually increases the asymmetry of the profile.
The Fano function arises from the coherent superposition of contributions
to the S matrix coming from different poles
and the linear function is determinate by the nonresonant channels
whose contributions add noncoherently. 
Therefore we have demonstrated that the asymmetry of the profile in
conductance does not necessarily involve the coupling between two
different channels as in the usual scenario to explain 
Fano resonances \cite{noeckel}.
In addition, the asymmetry of the conductance profile is not necessary
the asymmetry of the transmission peak.

In our theory of the resonant transport all parameter necessary to
describe the conductance profile can be calculated microscopically
and we can evaluate separately the coherent and the noncoherent 
parts of the conductance. The two contributions arise naturally
in our formalism of the coherent transport through open systems.
We do not exclude the existence of the incoherent processes in nanostructures
which can contribute to the conductance, but  we have fond that the
coherent processes contribute coherently and noncoherently 
to the conductance.

\subsection{CAPACITANCE OF QUANTUM MIS-TYPE HETEROSTRUCTURES}

Further, we analyze a quantum MIS (metal - insulator - semiconductor)-type
heterostructure. The considered AlAs/GaAs structure\cite{dolgopolov}
consists of a sequence of layers grown on a $GaAs$ bulk material
given by, first,
$n-GaAs$ layer as a back contact, second,
intrinsic $GaAs$ layer as a spacer, third,
a short period $Al_xGa_{1-x}As/GaAs$ superlattice as
a blocking barrier and
finally a metallization as a top gate.
\begin{figure}[h]
\begin{center}
\includegraphics[width=3.5in]{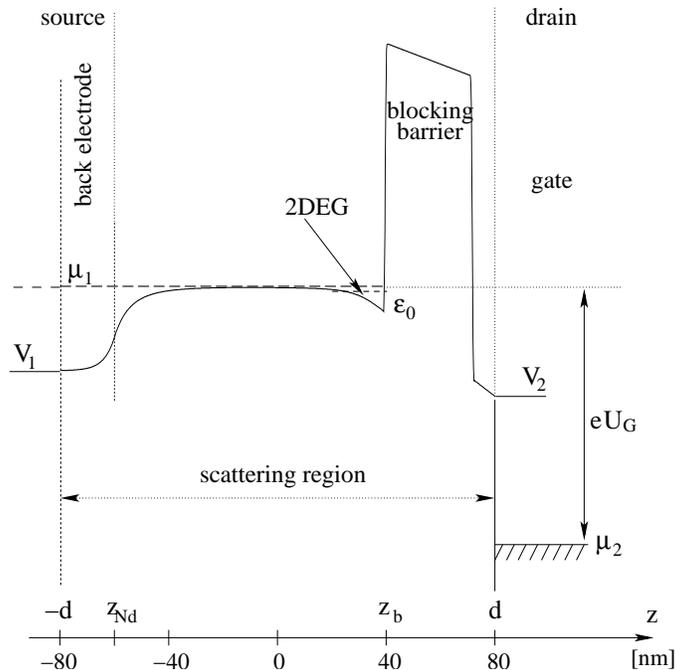}
\end{center}
\caption{Band diagram from self-consistent calculation
\cite{prb02,paul_cms}
for the MIS-type AlAs/GaAs heterostructure.
The structure parameters correspond to the experiments in 
Ref. \cite{dolgopolov}.}
\label{pot-H}
\end{figure}
The band structure, Fig. \ref{pot-H}, resulting from the
self-consistent calculations
in the Hartree approximation\cite{prb02,paul_cms}
shows for large enough positive gate voltages $U_G$
a potential well at the interface between the GaAs spacer
and the blocking
barrier, where a field induced two-dimensional electron gas (2DEG)
develops.

The blocking barrier is now assumed
to suppress charge
transfer completely and  we consider the
limit of small frequencies. 
Then,
the tunnel capacitor in Fig.\ \ref{pot-H}
becomes equivalent to a simple plate capacitor \cite{metal}.
The charge corresponding to the left plate is distributed
in the region $-d < z < z_b$ (see. Fig. \ref{pot-H})
and can be evaluated using Gauss' law,
$Q_1 = - S\kappa(\partial V/\partial z) (z=z_b)$.
Here
$z_b$ is the interface between the barrier and the $GaAs$-spacer layer,
$S$ is the area of the sample,
and $\kappa$ is the dielectric permittivity.
The top gate of the structure acts as the right plate of the
capacitor and the charge on it is $Q_2 = -Q_1$.
The capacitance is then readily found as
\begin{equation}
C = \left|\frac{\partial Q_1}{\partial U_g} \right| 
  = \left|\frac{\partial Q_2}{\partial U_g} \right|.
\label{capacitate}
\end{equation}

\begin{figure}[h]
\begin{center}
\noindent\includegraphics[width=4.5in]{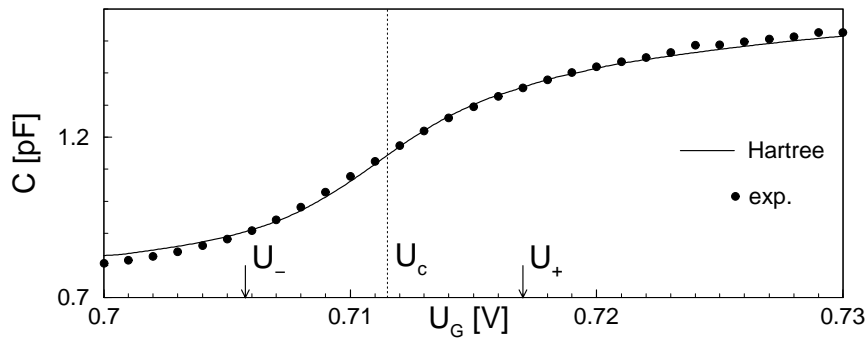}
\end{center}
\caption{
C-V curve, experimental data (filled circles)
\protect\cite{dolgopolov}
and numerical calculation  (solid line). }
\label{cap_H}
\end{figure}

In Fig.\ \ref{cap_H} we compare the  experimental C-V-curve
\cite{dolgopolov} with the results of our model \cite{prb02}.
For the numerical calculations, we use the parameters corresponding to the
experiments in Ref. \cite{dolgopolov}.
Because 
the work function of the metal contact is not precisely known
we shift the theoretical voltage scale $U_g$ 
with respect to the experimental one $U_G$ ($U_G=0.701V+U_g$), 
so that the centers of the steps coincide.
$U_g=0$ corresponds to the flat band configuration.
It can be seen that for such a system the C-V
curve takes the form of a broadened step
located between a low voltage, $U_g<U_-$,
and a high-voltage, $U_g>U_+$, plateau.
We define the gate voltage $U_c$ at the center of the step by 
$d^2C/dU_g^2=0$; $U_-$ and $U_+$ correspond
to the gate voltages where $|dC/dU_g|$
takes half  of its maximum value.

\begin{figure}[ht]
\begin{center}
\noindent\includegraphics[width=3.5in]{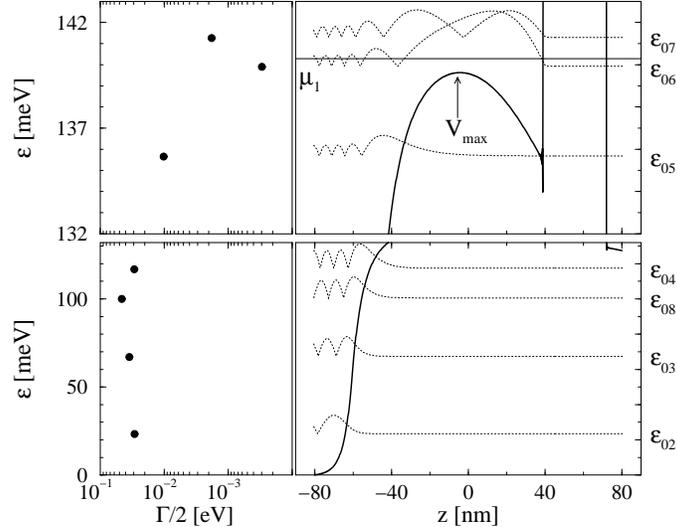}
\end{center}
\caption{Left:  Position of the resonance poles
in the complex energy plane determined by Eq.\ (\ref{pole1}).
Right: Potential energy $V(z)$ (solid line) for $U_g=16\;$mV
and $| \psi^{(1)}(\epsilon,z)|^2$ 
at the resonance energies (dotted line).
The upper plots present in detail an energy domain around $\mu_1$.
$V_{\max}$ marks the height of the
wide and shallow barrier which is formed between the back contact and
the blocking barrier,
$V_{\max}=\max_{-d \le z <  z_b}[V(z)]$.}
\label{fig_poles_016}
\end{figure}

\begin{figure}[ht]
\begin{center}
\noindent\includegraphics[width=3.85in]{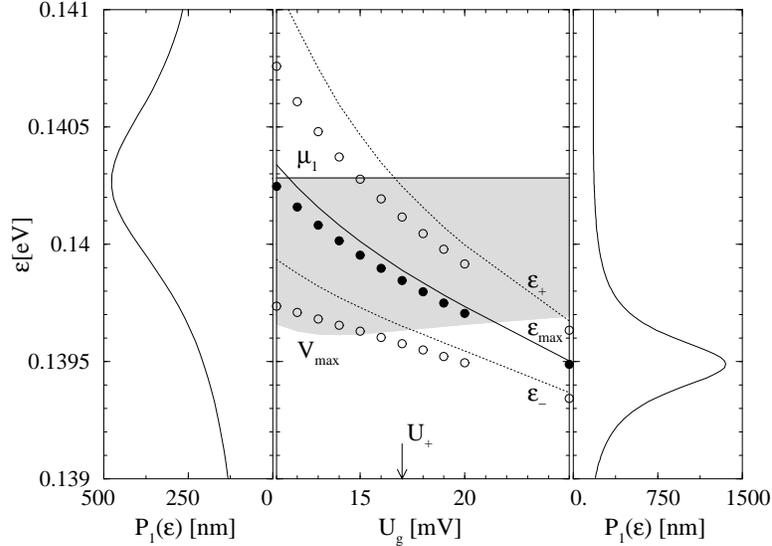}
\end{center}
\caption{Left part: Probability distribution $P_1(\epsilon)$
around the resonance energy $\epsilon_0$ for $U_g=11\;$mV.
Right side: Probability distribution $P_1(\epsilon)$
around the resonance energy $\epsilon_0$ for $U_g=25\;$mV.
Middle part:
The energy $\epsilon_{\max}$ of the maximum of $P_1(\epsilon)$
(solid line),
and the energies $\epsilon_{\pm}$ at which
$P_1(\epsilon)$ takes half the maximum value (dotted lines),
the resonant energies $\epsilon_0$ (filled circles)
and the energies $\epsilon_0 \pm \Gamma/2$ (circles).
Shaded area: energies which correspond to the classically allowed channel
($\mu_1 > \epsilon > V_{\max})$.
}
\label{fig_poles_016_2}
\end{figure}

To understand the variation of the capacitance with the
gate voltage in the step domain we 
perform 
the pole analysis. 
For the scattering potential of the biased structure,
Fig. \ref{fig_poles_016} illustrates
the pole energies with the real part
in the energy domain of the occupied states 
and the scattering functions at the resonance
energies corresponding to these poles.
This figure shows that there are two types of poles:
the first one is associated with the scattering states of
the electrons confined in the region close to the back contact.
The corresponding 
charge distribution varies very slowly with
the applied bias \cite{prb02,dolgopolov}.
Therefore, these resonances do not play any role in the
capacitance variation with $U_g$.
The second type of poles, with a smaller imaginary part,
corresponds to the states of the electrons
localized at the interface between the GaAs spacer layer
and the blocking barrier.
In our case there is a single pole
with the resonance energy below the chemical potential $\mu_1$ and
for which $| \psi^{(1)}(\epsilon,z)|^2$
has a maximum in the region of the potential quantum well \cite{prb02}.
We denote the complex energy of this pole by
$\bar{\epsilon}_0= \epsilon_0-i \Gamma/2$.
The imaginary part of this pole is smaller than that of the other poles
and smaller than the gap between two
adjacent resonance energies.
The charge accumulated in the quantum well  varies strongly with $U_g$
and, consequently, the changes of the capacitance
can be directly connected to the resonances of the second type.
A quantity which gives information about the contributions
of different poles to the charge accumulated in the system is
the probability distribution $P_s(\epsilon)$ 
[Eq.  \ref{Peps})]. In Fig. \ref{fig_poles_016_2},
$P_1(\epsilon)$
is plotted for two values of the applied bias: $U_g=11\;$meV,
corresponding approximatively to the center of the capacitance step
and  $U_g=25\;$meV, corresponding to the high voltage plateau of the 
capacitance curve (Fig. \ref{cap_H}). 
In the middle part of Fig. \ref{fig_poles_016_2}
the resonance energy $\epsilon_0$ and the energies 
$\epsilon_0 \pm \Gamma/2$ are presented as a function of $U_g$
together with the energy of the maximum of
$P_1(\epsilon)$, $\epsilon_{\max}$,
and the energies $\epsilon_{\pm}$ where
$P_1(\epsilon)$ takes half the maximum value.
For comparison the chemical potential and the height of the
wide and shallow barrier which is formed between the back contact and
the blocking barrier, 
$V_{\max}=\max_{-d \le z <  z_b}[V(z)]$, are also plotted.  
In this situation $V_{\max}$ is below  the chemical potential $\mu_1$
and a classically allowed channel for the electrons is opened.
For small positive values of $U_g$ the maximum of $P_1(\epsilon)$
lies in this channel.
With increasing $U_g$ a potential quantum well is formed at the
interface between the spacer layer and the blocking barrier
and $P_1(\epsilon)$
has a sharp maximum which
is located in an energy
range with no classically
allowed connection between the electrons in the quantum well
and the back contact.
So we can conclude that for positive values of $U_g$ $P_1(\epsilon)$
has a pronounced maximum centered on $\epsilon_0$
with a well defined width ($\simeq \Gamma$).
The charge accumulation in the field induced
quantum well of the quantum MIS type semiconductor 
is characterized mainly by the resonance associated with the pole
$\epsilon_0-i\Gamma/2$.
This resonance changes its character from an intermediate resonance
to a quasibound state.
For small applied biases ($U_- < U_g < U_+$)
this resonance has the character of an intermediate resonance
\cite{prb02}: i) its energy lies in the classically allowed range
(Fig. \ref{fig_poles_016_2}, middle part)
and, therefore,  is a Fabry-Perot type resonance;
ii) it is located in the space between the contact and the region where
an isolated 2DEG is formed at large values of $V_g$
(Fig. 3b.) Ref. \cite{prb02});
iii) its shape is strongly asymmetric
(Fig. \ref{fig_poles_016_2}, left part).
With increasing $U_g$ the intermediate resonance
turns into a quasibound state
which, in contrast to the intermediate resonance,
is connected to the back contact only via the tunneling effect.
The resonance line of $P_1(\epsilon)$ narrows and
tends to become symmetric (Fig. \ref{fig_poles_016_2}, right part).
The excellent quantitative  agreement between the experimental data and
the modeled capacitance curve, Fig. \ref{cap_H},
demonstrates that the electronic states, which are important for 
the capacitance step and
the subsequent high-voltage plateau are derived from a single resonance.

\section{CONCLUSIONS}
\label{conclusions}

We first analyze the electronic scattering states in
semiconductor nanostructures and, on this basis, the transport
properties of the electron gas: conductance and capacitance.
The considered systems are mesoscopic and the transport is 
supposed ballistic. 
The physics of these systems is often dominated by resonances 
which are broadened due to the open character of the systems.
We develop a theory of resonant transport in 
semiconductor nanostructures valid for all coupling regimes
between the quantum system and contacts, i.e. for open systems 
as well as for almost-closed quantum systems.
To separate weakly from strongly energy dependent contributions
on the scale of the width of the resonance we employ a representation 
of the {\bf S} matrix in terms of the {\bf R} matrix. We find a simple
procedure to determine the poles of the {\bf S} matrix 
in the complex energy plane, which gives the position  
and the width of the resonances. Numerical results show the remarkable
accuracy of our procedure which is applied to a great variety of resonances
types: quasibound states, Fowler-Nordheim, and Fabry-Perot resonances.
In case of separate resonances,
the linearization of the weakly energy dependent part in {\bf S} matrix
yields an analytical expression of the line shape 
which is a Fano profile with a complex asymmetry parameter.
The resonant theory of transport is applied to study 
the conductance through a quantum dot embedded in a quantum wire and 
the capacitance of a 2DEG formed at the interface between the spacer
layer and the blocking barrier in a MIS-type 
semiconductor heterostructure. 

\appendix
\section{ORTHOGONALITY OF THE SCATTERING FUNCTIONS}
\label{A-ortho}

The main aim of this section is to demonstrate Eq. (\ref{ortho}) 
using the expressions (\ref{psi-out-1}), (\ref{psi-out-2}),
and (\ref{psi-in}) of the scattering functions
 and the relation
(\ref{rel-StR}) between the {\bf S} and the {\bf R} matrix.
For an effective calculation of the scalar product between 
$\psi^{(s)}(\epsilon;z)$ and $\psi^{(s')}(\epsilon';z)$
it is necessary to split the integral
into two contributions, the first one given by the scattering domain 
and the second  one given by the lateral regions (outside the scattering
domain), 
\begin{equation}
\int_{-\infty}^{\infty} dz \left[\psi^{(s)}(\epsilon;z)\right]^*
                           \psi^{(s')}(\epsilon';z)
= I^{int}_{ss'}(\epsilon,\epsilon')
 +I^{ext}_{ss'}(\epsilon,\epsilon').
\label{ortho-det}
\end{equation} 
Further we are interested in evaluating the above integral
only in the cases in which the energies $\epsilon$ and $\epsilon'$ are in
the same domain: $\epsilon,\;\epsilon' > V_M$
or $V_m < \epsilon,\;\epsilon' < V_M$.
If the two energies belong to different domains, the corresponding
eigenfunctions are orthogonal due to the self-adjointness of the Hamilton
operator in Eq. (\ref{eq-S-1D}) and consequently Eq. (\ref{ortho})
is fulfilled.

On integrating over the lateral
regions such terms occur as
\begin{equation}
\int_a^{\infty} dz e^{i(k-k')z} = e^{i(k-k')a}
                                  \left[ i {\cal P}\frac{1}{k-k'}
                                        +\pi \delta(k-k')
                                  \right],
\label{int-aux}
\end{equation}                               
where ${\cal P}$ denotes the Cauchy principal part,
\begin{equation}
{\cal P}\frac{1}{k-k'} 
= \lim_{\mbox{\footnotesize $\begin{array}{c}
                             \eta \rightarrow 0 \\
                             \eta > 0                        
                             \end{array}$}}
       \frac{k-k'}{(k-k')^2+\eta^2},
\label{Cauchy-P}
\end{equation}
and after a laborious calculation we obtain
\begin{eqnarray}
I^{ext}_{ss'}(\epsilon,\epsilon')
& = & \theta(\epsilon-V_s) \theta(\epsilon'-V_{s'}) 
      \left\{\delta \left( k^*_1(\epsilon)-k_1(\epsilon') \right)
             \left[ \delta_{1s} \delta_{1s'}
                   +{\bf S}^*_{1s}(\epsilon) {\bf S}_{1s'}(\epsilon')
             \right]
      \right.
\nonumber \\
&   &        \hspace*{3cm}
            +\delta \left( k^*_1(\epsilon)+k_1(\epsilon') \right)
             \left[ \delta_{1s} {\bf S}_{1s'}(\epsilon')
                   +{\bf S}^*_{1s}(\epsilon) \delta_{1s'}
             \right]
\nonumber \\
&   &        \hspace*{3cm}
            +\delta \left( k^*_2(\epsilon)-k_2(\epsilon') \right)
             \left[ \delta_{2s} \delta_{2s'}
                   +{\bf S}^*_{2s}(\epsilon) {\bf S}_{2s'}(\epsilon')
             \right]
\nonumber \\
&   & \left. \hspace*{3cm}
            +\delta \left( k^*_2(\epsilon)+k_2(\epsilon') \right)
             \left[ \delta_{2s} {\bf S}_{2s'}(\epsilon')
                   +{\bf S}^*_{2s}(\epsilon) \delta_{2s'}
             \right]
      \right\}/2
\nonumber \\
&   &+{\cal P}\frac{1}{[k_1^*(\epsilon)]^2-k_1^2(\epsilon')} 
      \left[ \left(\Psiv_S(\epsilon)\right)_{1s}^*
             \left(\Psiv(\epsilon')\right)_{1s'}
            -\left(\Psiv(\epsilon)\right)_{1s}^*
             \left(\Psiv_S(\epsilon')\right)_{1s'}
      \right]
\nonumber \\
&   &+{\cal P} \frac{1}{[k_2^*(\epsilon)]^*-k_2^2(\epsilon')}
      \left[ \left(\Psiv_S(\epsilon)\right)_{2s}^*
             \left(\Psiv(\epsilon')\right)_{2s'}
            -\left(\Psiv(\epsilon)\right)_{2s}^*
             \left(\Psiv_S(\epsilon')\right)_{2s'}
      \right].
\label{I-out-1}
\end{eqnarray}
To evaluate the delta functions and their coefficients 
we have to discuss the cases
$\epsilon,\;\epsilon' > V_M$
and $V_m < \epsilon,\;\epsilon' < V_M$ separately.
For $\epsilon,\;\epsilon' > V_M$
we obtain
$\theta(\epsilon-V_s) \theta(\epsilon'-V_{s'})=1$,
$\delta(k^*_s(\epsilon)-k_s(\epsilon'))
 =\delta(\epsilon-\epsilon')/g_s(\epsilon)$ with $g_s(\epsilon)$
defined by Eq. (\ref{gs}), $s=1,2$ and 
$\delta(k^*_s(\epsilon)+k_s(\epsilon'))=0$.
Using the relation between the ${\bf S}$- and 
the $\tilde{{\bf S}}$ matrix and the unitarity of $\tilde{{\bf S}}$
[Eq. (\ref{unitarity})] the first term in the expression (\ref{I-out-1})
becomes $\delta(\epsilon-\epsilon')\delta_{ss'}/g_s(\epsilon)$. 
In the case $V_m < \epsilon,\;\epsilon' < V_M$
it is found that
$\theta(\epsilon-V_s) \theta(\epsilon'-V_{s'})=\delta_{ss'} \delta_{sm}$,
$\delta(k^*_m(\epsilon)-k_m(\epsilon'))
     =\delta(\epsilon-\epsilon')/g_m(\epsilon)$ with $g_m(\epsilon)$
defined by Eq. (\ref{gs}) and 
$\delta(k^*_m(\epsilon)+k_m(\epsilon'))=0$.
Taking also into account that for considered energies
${\bf S}_{mm}(\epsilon)=1$, we obtain the same result as in the first
case.
In the last two terms of Eq. (\ref{I-out-1})
we substitute $[k^*_s(\epsilon)]^2-k_s^2(\epsilon')
           =(\epsilon-\epsilon') 2 m^*/\hbar^2$,
$s=1,2$, which directly follows from the definition (\ref{k}) of $k_s$
and
$I^{ext}_{ss'}(\epsilon,\epsilon')$ becomes
\begin{equation}
I^{ext}_{ss'}(\epsilon,\epsilon')
= \frac{\theta(\epsilon-V_s) 
        \delta \left( \epsilon-\epsilon' \right) \delta_{ss'}}
       {g_s(\epsilon)}
 -{\bf M}_{ss'}(\epsilon,\epsilon'),
\label{I-out-2}
\end{equation}
where the matrix ${\bf M}(\epsilon,\epsilon')$ is defined as
\begin{equation}
{\bf M}(\epsilon,\epsilon') 
= \frac{\hbar^2}{2 m^*} 
  \frac{1}{\epsilon-\epsilon'}
  \left[ \Psiv^\dagger(\epsilon) \Psiv_S(\epsilon')
        -\Psiv^\dagger_S(\epsilon) \Psiv(\epsilon')
  \right]
\label{rho-mat}
\end{equation}
for  $\epsilon \ne \epsilon'$ and
\begin{equation}
{\bf M}(\epsilon,\epsilon) 
= \frac{\hbar^2}{2 m^*}
  \left[ \Psiv^\dagger_S(\epsilon) \frac{d \Psiv}{d \epsilon}
        -\Psiv^\dagger(\epsilon) \frac{d \Psiv_S}{d \epsilon}
  \right]
\label{rho-mat-ee}
\end{equation}
for $\epsilon=\epsilon'$.
 
To calculate $I_{ss'}^{int}(\epsilon,\epsilon')$ we use the expression 
(\ref{psi-in}) of the scattering functions
inside the scattering region and find
\begin{eqnarray}
I_{ss'}^{int}(\epsilon,\epsilon')
& = & \sum_{i,j=1,2} 
         \left(\frac{2d}{\pi} \right)^2
         \left( \Psiv_S^\dagger(\epsilon) \right)_{si}
         \left( \Psiv_S(\epsilon') \right)_{js'}
         \int_{-d}^{d} dz R(\epsilon;(-1)^i d,z) R(\epsilon';(-1)^j d,z).
\label{I-int-1}
\end{eqnarray}       
The integral in the above expression of
$I_{ss'}^{int}(\epsilon,\epsilon')$
can be easily calculated if the R functions are replaced by their definition
(\ref{R-def}) and the orthogonality condition of the Wigner-Eisenbud
functions [Eq. (\ref{WE-ortho})] is used. After that a very simple trick,
\begin{equation}
\frac{1}{(\epsilon-\epsilon_l)(\epsilon'-\epsilon_l)}
={\cal P}\frac{1}{\epsilon-\epsilon'}
 \left[ \frac{1}{\epsilon'-\epsilon_l} 
       -\frac{1}{\epsilon-\epsilon_l}
 \right],
\end{equation}
where the Cauchy principal part is defined by Eq. (\ref{Cauchy-P}),
allows us to write
\begin{eqnarray}
\int_{-d}^{d} dz R(\epsilon;(-1)^i d,z) R(\epsilon';(-1)^j d,z)
\nonumber \\
\qquad \qquad 
= \frac{\hbar^2}{2m^*} \frac{\pi}{2d}
  {\cal P}\frac{1}{\epsilon-\epsilon'}
  \left[ R(\epsilon'; (-1)^i d, (-1)^j d)
        -R(\epsilon; (-1)^i d, (-1)^j d)
  \right].
\label{int-RR}
\end{eqnarray}
Inserting Eq. (\ref{int-RR}) in Eq. (\ref{I-int-1})
we obtain
\begin{equation}
I_{ss'}^{int}(\epsilon,\epsilon')
= {\bf M}_{ss'}(\epsilon,\epsilon').
\label{I-int-2}
\end{equation}
Using the expressions of the integrals inside and outside the scattering
region [Eqs. (\ref{I-int-2}) and (\ref{I-out-2}), respectively]
the orthogonality condition [Eq. (\ref{ortho})] of the scattering functions
follows from Eq. (\ref{ortho-det}).


\begin{thebibliography}{100}
\addcontentsline{toc}{chapter}{Bibliography}
\bibitem{ferry}
Ferry D. K.and Goodnick S. M. 1999, Transport in Nanostructures,
Cambridge University Press, Cambridge.
\bibitem{kukulin}
Kukulin V. I., Krasnopolsky V. M., and Horacek J. 1989,
Theory of Resonances - Principles and Applications,
Kluwer Academic Publishers, Dordrecht.
\bibitem{esaki1}
Esaki, L. and Tsu R. 1970, IBM J. Res. Develop. {\bf 14}, 61.
\bibitem{esaki2}
Tsu R., and Esaki L. 1973, Appl. Phys. Lett. {\bf 22}, 562.
\bibitem{kelly}
Kelly M.J. 1995, Low-dimensional Semiconductors, 
Clarendon Press, Oxford.
\bibitem{ando}
Ando T., Arakawa Y., Furuya K., Komiyama S., and Nakashima H.(Eds.),
1998,Mesoscopic Physics and Electronics, Springer Verlag, Berlin.
\bibitem{univ} 
Fowler A. B., Hartstein A., and Webb R. A. 1982,
Phys. Rev. Lett. {\bf 48}, 196;
\bibitem{ahar}
B\"uttiker M., Imry Y., Landauer R., and Pinhas S. 1985,
Phys. Rev. B {\bf 31}, 6207.
\bibitem{quanha}
B\"uttiker M. 1988,
Phys. Rev. B {\bf 38}, 9375.
\bibitem{quapoi}
van Wees B. J. et al. 1988,
Phys. Rev. Lett. {\bf 60}, 848;
Szafer A. and Stone A.D. 1989,
Phys. Rev. Lett. {\bf 62}, 300.
\bibitem{coul}
Meirav U., Kastner M. A., and Wind S. J. 1990,
Phys. Rev. Lett.  {\bf 65}, 771.
\bibitem{chao}
Jalabert R. A., Stone A. D., and Alhassid Y. 1992,
Phys. Rev. Lett. {\bf 68}, 3468.
\bibitem{kondo}
Goldhaber-Gordon D., Shtrikman H., Mahalu D., Abusch-Magder D.,
Meirav U., and Kastner M. A. 1998, 
Nature, {\bf 391}, 156.
\bibitem{lanbue}
Landauer R. 1957, IBM J. Res. Develop. {\bf 1}, 223;
Landauer R. 1987, Z. Phys. {\bf B 68}, 217;
Fisher D. S. and Lee P. A. 1981, Phys. Rev. B {\bf 23}, 6851;
B\"uttiker M. 1986, Phys. Rev. Lett. {\bf 57}, 1761;
B\"uttiker M. 1988, IBM J. Res. Develop. {\bf 32}, 317;
Stone A. D. and Szafer A. 1988, IBM J. Res. Develop. {\bf 32}, 384.
\bibitem{bue92}
B\"uttiker M. 1992, Phys. Rev. B {\bf 46}, 12485.
\bibitem{nucr}
Humblet J. and Rosenfeld L. 1961, Nucl. Phys. {\bf 26}, 529.
\bibitem{goeres}
G\"ores J., Goldhaber-Gordon D., Heemeyer S. , and Kastner M.A.,
Shtrikman H., Mahalu D., and Meirav U. 2000,
Phys. Rev. B {\bf 62}, 2188.
\bibitem{schmidt}
Schmidt T., K\"onig P., McCann E., Fal'ko V. I., and Haug R. J. 2001,
Phys. Rev. Lett. {\bf 86 }, 276.
\bibitem{lal}
Lal S., Rao S., and Sen D. 2001, Rhys. Rev. Lett. {\bf 87}, 026801.
\bibitem{bastard}
Bastard G., Brum J. A., and Ferreira R. 1991, 
Sol.State Phys. {\bf 44}, 229.
\bibitem{taylor}
Taylor J. R. 1972, Scattering Theory: The Quantum Theory of
Nonrelativistic Collisions, John Wiley \& Sons, Inc., New York. 
\bibitem{WE}
Wigner E. P.and Eisenbud L. 1947,
Phys. Rev. {\bf 72}, 29.
\bibitem{lane}
Lane A. M. and Thomas R. G. 1958,
Rev. Mod. Phys. {\bf 30}, 257.
\bibitem{Rmat-transport}
Smr\v{c}ka L. 1990 ,
Superlattices Microstruct. {\bf 8}, 221;
Wulf U., Ku\v{c}era J., Racec P. N., and Sigmund E. 1998,
Phys. Rev. B {\bf 58}, 16209;
Alhassid Y. 2000, 
Rev. Mod. Phys. {\bf 72}, 895.
\bibitem{thesis}
Racec E. R. 2002, Ph.D. Thesis, University of Technology Cottbus.
\bibitem{huang}
Huang K. 1965, Statistical Mechanics, John Wiley \& Sons, Inc., New York.
\bibitem{bue85}
B\"uttiker M., Imry Y., Landauer R., and Pinhas S. 1985,
Phys. Rev. B {\bf 31}, 6207.
\bibitem{prb01}
Racec  E. R. and Wulf U. 2001, 
Phys. Rev. B {\bf 64}, 115318.
\bibitem{sse}
Racec P. N., Wulf U. and Ku\v{ce}ra J. 2000, 
Solid-State Electronics 44, 881. 
\bibitem{prb02}
Racec P. N., Racec E. R., and Wulf U. 2002,
Phys. Rev. B 65, 193314.
\bibitem{fyodorov}
Fyodorov Y. V. and Sommers H. J. 1997,
J. Math. Phys. {\bf 38}, 1918.
\bibitem{bohm}
Bohm A. 1993, Quantum Mechanics, Springer, New York,
Chap. 18 Resonance Phenomena.
\bibitem{kampen}
van Kampen N. G. 1953, Phys. Rev. {\bf 91}, 1267. 
\bibitem{noeckel}
N\"ockel J. U. and Stone A. D. 1995,
Phys.Rev. B {\bf 51}, 17219;
N\"ockel J. U. and Stone A. D. 1994,
Phys. Rev. B {\bf 50} 17415.
\bibitem{bue88}
B\"uttiker M. 1988,
Phys. Rev. B {\bf 38}, 12724.
\bibitem{landau}
Landau L. D. and  Lifschitz  E. M. 1977, 
Quantum Mechanics (Non-Relativistic Theory), Pergamon, Oxford.
\bibitem{WB}
Alhassid Y. and Attias H. 1996,
Phys. Rev. B {\bf 54}, 2696;
Alhassid Y. and H. Attias H. 1996,
Phys. Rev. Lett. {\bf 76}, 1711.
\bibitem{lat}
Kastner M. A. 1993, Phys. Today {\bf 46}, 24;
\bibitem{tarucha95}
Tarucha, S., Austing D. G., Honda T., van der Hage R. J., and
Kouwenhoven L. P. 1996,
Phys. Rev. Lett. {\bf 77}, 3613.
\bibitem{paul_cms}
Racec P. N., Racec E. R., and Wulf U. 2001,
Computational Materials Science {\bf 21}, 475.
\bibitem{paul_phd}
Racec P. N. 2002, Ph.D. Thesis, University of Technology Cottbus.
\bibitem{dolgopolov} 
Dolgopolov V. T., Shashkin A. A., Aristov A. V.,
Schmerek D., Drexler H., Hansen W., Kotthaus J. P., and Holland M. 1996,
Phys. Low-Dim. Struct. {\bf 6}, 1.
\bibitem{metal}
Cavicchi R. E. and Silsbee R.H. 1988,
Phys. Rev. B {\bf 37}, 706.
\end{thebibliography}
\end{document}